\documentstyle[12pt,mystyle,epsf]{article}

\epsfverbosetrue

 1
 1
 1

\newcommand{\wti}{\widetilde}

\newcommand{\tb}{\tan \beta}
\newcommand{\bsgamma}{b\to s \gamma}
\newcommand{\nn}{\nonumber}
\newcommand{\ggp}{{g^2 \!+ g^{\prime 2}}}
\newcommand{\ggm}{{g^2 \!- g^{\prime 2}}}
\newcommand{\ale}{{\alpha}_{\rm em}}

\newcommand{\go}[1]{\gamma^{#1}}
\newcommand{\gu}[1]{\gamma_{#1}}

\def\as{\alpha_{\rm S}}
\def\gs{g_{\rm S}}

\def\GeV{\,{\rm GeV}}
\def\MSSM{{\rm MSSM}}
\def\BR{{B\!R}}

\def\ltap{\ \raisebox{-.4ex}{\rlap{$\sim$}} \raisebox{.4ex}{$<$}\ }
\def\gtap{\ \raisebox{-.4ex}{\rlap{$\sim$}} \raisebox{.4ex}{$>$}\ }
\begin{document}
%
%

\begin{titlepage}
\noindent
DESY 93-090  \hfill  ISSN 0418-9833  \\
August 93\hfill                        \\[9ex]
\begin{center}
{\Large \bf The decay $b \to s \gamma$ in the MSSM revisited} \\[11ex]

{\large F.M.\ Borzumati                             }    \\[1.5ex]
{\it II.\ Institut f\"ur Theoretische Physik
\footnote{Supported by the Bundesministerium f\"ur Forschung und
 Technologie, 05 5HH 91P(8), Bonn, FRG.   } }       \\
{\it Universit\"at Hamburg, 22761 Hamburg, Germany}      \\[21ex]

{\large \bf Abstract}
\end{center}
\begin{quotation}
We present a re-analysis of the decay $\bsgamma$ in the
Minimal Supersymmetric Standard Model with radiatively induced
breaking of $SU(2) \times U(1)$. We extend this analysis to regions
of the supersymmetric parameter space wider than those previously
studied. Results are explicitly presented for $m_t=150\GeV$. We
emphasize the consequences for future searches of charged Higgs and
charginos from a measurement of the branching ratio for this decay
compatible with the Standard Model prediction. In spite of the strong
sensitivity of this decay to the effects of supersymmetry, we find
that, at the moment, no lower limit on the mass of these particles
can be obtained. The large chargino contributions to the branching
ratio for $\bsgamma$, for large values of $\tb$, are
effectively reduced by pushing the lightest eigenvalue of the
up-squark mass matrix closer to the electroweak scale, i.e. by
increasing the degree of degeneracy among the up-squarks.
\end{quotation}
\end{titlepage}

\section{Introduction and Motivations}

The recent results from the CLEO Collaboration \cite{CLEO} on
radiative decays of the $B$ meson have
rekindled
attention on the transition $\bsgamma$. Interesting is the improvement
of the upper limit on the branching ratio of this inclusive decay,
$\BR(\bsgamma)< 5.4\times 10^{-4} \,@\,95\%$\,CL, now closer to the
range of values allowed in the Standard Model (SM).

The possible implications of this measurement on the determination of
unknown parameters of the SM, however, seem nowadays much weaker than
they appeared in the past. The hope of pinning down the mass of the
top quark, $m_t$, from the inclusive process lost ground as
soon as it became clear how important a role was played by the QCD
corrections in the $\BR(\bsgamma)$~\cite{QCD,GRSPWI,OTHERS,GRICHO,MISIAK}.
The size of these corrections at the leading order (LO) in QCD brings
in a rather severe uncertainty on the theoretical determination of
this branching ratio, which we denote in the SM as
$\BR(\bsgamma)\vert_{\rm SM}$. This uncertainty, primarily due to the
unknown value of the scale $\mu$ at which the strong coupling constant
$\alpha_{\rm s}(\mu)$ should be evaluated, amounts roughly to a factor
of two~\cite{ALIGR}. It has been recently observed that the only new
outcome of the measurement of $\BR(\bsgamma)\vert_{\rm SM}$ maybe
the determination of the value of the Cabibbo-Kobayashi-Maskawa (CKM)
matrix element $K_{\rm ts}$ \cite{ALIGR}. The predictiveness of such a
measurement will clearly improve once the next-to-leading order (NLO) QCD
corrections to the SM effective Hamiltonian are made available.

The poor knowledge of $\BR(\bsgamma)\vert_{\rm SM}$ cannot be overlooked
when one attempts to use this decay to determine the value of parameters
present in extensions of the SM. It is undoubted, by now, that this
decay can be quite sensitive to the presence of particles and
interactions predicted in models which enlarge the SM Hamiltonian.
Nevertheless, given the imminent detection of $\bsgamma$,
the question one should try to answer, as precisely as possible, is
to what extent relevant portions of the parameter spaces
of these models can actually be excluded.

Several papers have recently appeared rediscussing the implications of
this decay for supersymmetric models~%
\cite{HEWETT,BARGER,DIAZ,OSHIMO,TANIMOTO,BARGIUD,NANOPOU}. In
particular, it can be inferred from references \cite{HEWETT} and
\cite{BARGER} that the present CLEO upper limit, puts already
strong constraints on the mass of the supersymmetric charged Higgs,
$m_{H^\pm}$, independently of the value of $\tb$, the ratio
of the two vacuum expectation values, $v_1$,\,$v_2$, present in
this model ($\tb=v_2/v_1$). These constraints would almost
pre-empt the charged Higgs searches at LEPII, close the possibility
of the top quark decay $t\to bH^+$ and the region of parameter space
$(\tb,m_{H^\pm})$ inaccessible at SSC/LHC. The conclusions
reached in these papers rely on the assumption that the charged Higgs
contribution to the one-loop diagrams mediating the $\bsgamma$ decay
is the dominant one among all the supersymmetric contributions.

Five different sets of contributions to the decay
$\bsgamma$ are present in supersymmetry. They can be classified
according to the virtual particles exchanged in the loop: {\it a)} the SM
contribution with exchange of $W^-$ and up-quarks; {\it b)} the charged
Higgs contribution with $H^-$ and up-quarks; {\it c)} the chargino
contribution with $\widetilde\chi^-$ and up-squarks ($\wti u$);
{\it d)} the gluino contribution with $\tilde g$ and down-squarks
($\wti d$); and finally {\it e)} the neutralino contribution
with $\widetilde\chi^0$ and down-squarks.
A complete evaluation of these contributions
was performed in ref.~\cite{US} within the framework of the
Minimal Supersymmetric Standard Model (MSSM) with
radiatively induced breaking of $SU(2)\times U(1)$.

The MSSM is the low-energy remnant of spontaneously
broken $N=1$ supergravity theories with: i) canonical kinetic terms
(flat K\"ahler metric) for all the scalar fields;
ii) no extra superfields besides
those needed for a minimal supersymmetric version of the SM; iii) no
baryon- and/or lepton-number violating terms in the superpotential.
These restrictions lead to a simplified structure of the soft
supersymmetry-breaking terms such that the overall number of
parameters of the MSSM, in addition to the gauge and yukawa couplings,
can be reduced to {\it five}. They are:
i) $\mu$, the dimensional parameter in the superpotential which couples
the superfields containing the two Higgs doublets;
ii) $m$, the common soft breaking mass for all the scalars of the
theory; iii) $M$, the common soft breaking gaugino mass;
iv) $A$, and v) $B$, the dimensionless parameters appearing
respectively in the trilinear ad bilinear
soft breaking scalar terms obtained by taking the scalar components of
the corresponding terms in the superpotential.
In the presence of a flat K\"ahler metric, $B$ and $A$ are connected
through the relation $B=A-1$.
The request of having the breaking of the gauge group
$SU(2) \times U(1)$ induced by renormalization effects enforces
functional relations among these initial parameters and the scale of
the electroweak breaking. Thus, the number of independent
parameters, in addition to those present in the SM, is
only {\it three}. We choose
them to be, as in~\cite{US}, $m,M,\tb$.

The entire low-energy spectrum and all the couplings
present in this model can be expressed in terms of these
parameters. The calculation of the individual
contributions to the total amplitude
${\cal A}(\bsgamma)\vert_\MSSM$ and the final branching ratio
$\BR(\bsgamma)\vert_\MSSM$ is then relatively straightforward.
Results of this calculation were presented in \cite{US}
only for the case of $m_t = 130\GeV$ and $\tb = 2,8$.

It was observed there that the two main supersymmetric contributions to
$\BR(\bsgamma)\vert_\MSSM$, competitive with the SM exchange of $W^-$
and top quark, were {\it b)} and {\it c)}. Smaller was found the
contribution {\it d)}, which had been cherished in
the past as the most likely mechanism to produce supersymmetric
enhancements of flavour changing neutral current (FCNC) processes.
The contribution {\it e)} appeared to be totally negligible. It was also
noticed that, in the two cases studied, the largest value of
$\BR(\bsgamma)\vert_\MSSM$ was obtained in a region of parameter
space were the Higgs contribution {\it b)} was, indeed, the dominant one.

Destructive interferences among the abovementioned classes of
contributions to the amplitude ${\cal A}(\bsgamma)\vert_\MSSM$ were
observed in non-trivial portions of the supersymmetric parameter
space. Nevertheless,
the total results obtained for $\BR(\bsgamma)\vert_\MSSM$ were
almost always showing
an effective enhancement of the SM prediction. Small suppressions were
observed, for $\tb=2$, only in a tiny region at the higher end
of the range of values considered for the
parameter $m$.
Finally, the band of possible values obtained
for $\BR(\bsgamma)\vert_\MSSM$, in the
case $\tb=8$, had a
width,
due to the variation of the remaining free parameter $M$,
which was fairly constant for increasing $m$.
Branching ratios exceeding the SM prediction
by more than a factor two could still be observed at the higher
end of $m$, where the Higgs contribution is rather small.

These last two occurrences can be easily explained if one neglects
for a moment the contributions {\it d)} and {\it e)}, which played a
minor role
in the two cases studied in \cite{US}. Since the Higgs contribution to
${\cal A}(\bsgamma)\vert_\MSSM$ has always the same sign of the
SM contribution, it is clear that
the small suppressions of
$\BR(\bsgamma)\vert_\MSSM$ with respect to the SM prediction
observed for $\tb = 2$, and the still sizable enhancement
present at large values of $m$ for $\tb = 8$ are due to
negative and positive chargino contributions in absolute size
bigger than the Higgs contribution.

These observations undermine already the main assumption of
refs.~\cite{HEWETT,BARGER}. The dominance of one of the supersymmetric
contributions over the remaining ones depend strongly on
the region of parameter space considered. Therefore, in general, the
burden of restrictions which experimental findings may impose on
the allowed values of $\BR(\bsgamma)\vert_\MSSM$ has to be
shared by all the supersymmetric
contributions. The question of where and if these restrictions can
be translated in
clear-cut bounds on specific masses remains still open.

Already in ref.~\cite{DIAZ}, where only the Higgs contribution
to $\bsgamma$ was considered, it was shown that a consistent
inclusion of radiative corrections to the tree level Higgs potential
brings back in the calculation of the branching ratio the dependence of
supersymmetric masses and couplings and
weakens the strong predictions of
\cite{HEWETT,BARGER}, at least for large enough values of
$\tb$.

The possibility of (positive and negative) chargino contributions sizably
exceeding, in absolute value, not only the Higgs contribution, but also
the SM one, for relatively large values of $\tb$,
$\tb \gtap 10$, was recently emphasized in
ref.~\cite{OSHIMO}. It was also shown that, always for
$\tb \gtap 10$, the gluino contribution may not be so
small, leading therefore to
further enhancements of $\BR(\bsgamma)\vert_{\rm MSSM}$.

Aim of this paper is to extend the analysis of ref.~\cite{US} to regions
of parameter space wide enough to allow: i) significant answers to the
questions raised in \cite{HEWETT,BARGER}; ii) confirmations of the
results obtained in~\cite{OSHIMO}; iii) an investigation of the
possibility that these
results may threaten the chargino searches
at LEP~II. Two well distinguished aspects
enter in the calculation of $\BR(\bsgamma)\vert_\MSSM$:
the calculation of the supersymmetric mass spectrum to be
inputted in $\BR(\bsgamma)\vert_\MSSM$ and
the actual calculation of the branching ratio. For details on these two
aspects, the reader is referred to
refs.~\cite{US,IO} and~\cite{US}, respectively. Nevertheless a few points
for each of them are reported and emphasized in Sects.~2 and~3, in order to
set the notation, to clarify approximations and assumptions made in this
analysis. Furthermore, some
features of the supersymmetric parameter space, relevant for the
$\bsgamma$ decay, are described in Sect.~3. A thorough discussion
of the results obtained is given in Sect.~4, followed then by
the conclusions. Finally, a list of misprints/errors for
refs.~\cite{US,IO} is given in the Appendix.

\newpage

\section{The Supersymmetric Mass Spectrum}
%
\subsection{Notation, Procedure and Inputs}

We briefly outline in this section the procedure used in the
calculation of the supersymmetric mass spectrum to be inputted in
$\BR(\bsgamma)\vert_\MSSM$. This calculation is performed
within the framework of the MSSM with spontaneous breaking
of the gauge group $SU(2)\times U(1)$ induced by renormalization effects.

At the electroweak scale, the supersymmetric mass spectrum is described
by a lagrangian which we denote as ${\cal{L}}_\MSSM(M_Z)$. As for the
calculation of $\BR(\bsgamma)\vert_\MSSM$, this is nothing more than a
SM lagrangian extended by the introduction of new fermion and scalar
particles and of new renormalizable interaction terms.

In particular, two scalars are present for each up- and
down-quark,
the up- and down-squarks
${\wti u}_i$,~${\wti d}_i$. Two and one scalars are also associated to
each charged and neutral lepton
the charged sleptons and sneutrinos ${\wti l}_i$,~${\wti \nu}_j$,
respectively. An additional charged scalar Higgs, $H^\pm$ and three neutral
Higgs, $H_1^0$,~$H_2^0$ (CP-even), and $H_3^0$ (CP-odd) are
present. Finally, one has two additional charged fermions, the
charginos ${\wti \chi_k}^-$, related to the supersymmetric partners of
the $W$-boson and of the charged higgs bosons, (the $W$-ino and
Higgs-ino, respectively) and four neutral fermions, the neutralinos
${\wti \chi_l}^0$, related to the supersymmetric partners of the three
neutral higgs and of the neutral gauge boson. The lightest of all these
sets of particles is usually labelled by $i\!=\!j\!=\!l\!=\!1$, except
for
the lighter chargino and the lightest neutral higgs,
conventionally denoted as ${\wti \chi_2}^-$ and $H_2^0$.

The complexity of ${\cal{L}}_\MSSM(M_Z)$ reduces considerably if one
relates it to the simpler form which ${\cal L}_\MSSM$ has at a scale
of the order of the Planck mass, ${\cal {L}}_\MSSM(M_P)$. For the
specific form of this lagrangian the reader is referred
to~\cite{US,IO}. We shall remind here only that ${\cal {L}}_\MSSM(M_P)$,
remnant of a theory with spontaneously broken $N=1$ local supersymmetry
after decoupling of gravity, consist of two terms: the globally
supersymmetric version of
the SM with two Higgs doublets (2HDM)~\footnote{Two types of
 two-Higgs-doublet-models are known, the so-called type~I and type~II,
 with different couplings of the Higgs bosons to fermions. By the
 acronym 2HDM we indicate throughout all this paper the type~II model,
 with Higgs bosons coupling to fermions as in the MSSM%
.},
which we denote as
${\cal {L}}^{^{\ {\rm gl-susy}}}_{\rm 2HDM}(M_P)$,
plus a collection of soft
terms, explicitly breaking this global supersymmetry,
${\cal {L}}_{\rm soft}(M_P)$.

${\cal {L}}^{^{\ {\rm gl-susy}}}_{\rm 2HDM}(M_P)$
has the minimal particles content (listed above) and minimal set of
interaction terms needed to supersymmetrize the SM. The presence of an
extra Higgs doublet,
with the corresponding supersymmetric partner,
has already been mentioned.
An extra bilinear term, with dimensional coupling $\mu$, mixes
the two Higgs superfields in the superpotential. No baryon-
and lepton-violating terms, are included in the superpotential.
Finally, canonical kinetic terms are assumed for all scalar
superfields in the
initial supergravity lagrangian.

The form of the soft breaking terms
${\cal {L}}_{\rm soft}(M_P)$ is a compromise between {\it a)} the
need of a structure general enough to account for different
possible mechanisms (unknown) of spontaneous breaking of the
underlying local supersymmetry, {\it b)} a criterion of simplicity
requiring the introduction of a minimum number of additional
parameters. All the scalar and gaugino mass terms in
${\cal {L}}_{\rm soft}(M_P)$ are
given the same couplings,
$m^2$ and $M$, respectively, and the trilinear and bilinear
soft-breaking terms, obtained by taking the scalar components of the
corresponding terms in the superpotential, appear with the two
dimensionless parameters $A$ and $B$.
The abovementioned requirement of canonical kinetic terms for all the
scalar fields in the underlying theory (flat K\"ahler metric) relates
$A$ and $B$ according to $B=A-1$.

These properties define what is known as MSSM. {\it Four}
is the number of new parameters of this model, in addition to
those present in the SM: $\mu$,~$m$,~$M$,~$A$.

As mentioned, the lagrangian ${\cal {L}}_\MSSM(M_P)$ describes an
effective model obtained after decoupling of all
the higher dimension operators weighted by
inverse powers of $M_P$,
in principle not negligible at scales ${\cal O}(M_P)$.
This model is embedded in a grand-unified theory,
with unification scale $M_X$, about three order of magnitude
below $M_P$. The presence of additional heavy particle associated
to the grand-unified gauge group should be kept into account when
evolving the parameters of this model to lower scales.
The procedure usually adopted is to neglect both type of terms.
All other renormalization effects between $M_P$ and $M_X$
are also ignored
and ${\cal L}_\MSSM(M_P)$ is simply equated to ${\cal L}_\MSSM(M_X)$.
A quantitative estimate of how these effects, if taken into account,
could alter the
rigid universality of couplings in ${\cal {L}}_\MSSM(M_X)$
and therefore affect the low-energy spectrum does not
exist~\footnote{There exist estimates of the uncertainties
 introduced  on the gauge couplings unification~\cite{LANGPOL}
 by non-renormalizable terms and threshold effects for the
 heavy particles. These are found to be small.}.

We follow here, as in~\cite{US}, this conventional procedure
and obtain the low-energy spectrum of the supersymmetric particles
listed before, as solution of a set of renormalization group~(RG)
equations. The initial conditions for these equations
can be easily expressed in terms of $\mu,m,M,A$
and the values of gauge and yukawa couplings at $M_X$.
The further request that the mass parameters for the scalar particles
evolve down to the electroweak scale in such a way to induce the
breaking of $SU(2) \times U(1)$ enforces functional relations among
the four parameters of the model and
allows to eliminate some of them.

Given the complexity of the full low-energy scalar
potential, one starts imposing that the correct vacuum is 
recovered through the minimization of a subset of it, i.e.
the Higgs potential, which at low energy reads:
\bea
\lefteqn{V \ = \
   \mu^2_1 \vert H_1\vert^2 + \mu^2_2 \vert H_2\vert^2
 - \mu^2_3 \left(\epsilon_{ij} H^i_1 H^j_2+ h.c.\right)
                                                       } \nonumber \\
 & &
 + \frac{1}{8} (\ggp)
          \left(\vert H_1\vert^2 \!+\vert H_2\vert^2 \right)
 + \frac{1}{4} (\ggm)  \vert H_1\vert^2  \vert H_2\vert^2
 - \frac{1}{2} g^2 \vert \epsilon_{ij} H^i_1 H^j_2 \vert^2\,,
\label{higgspot}
\eea
where $H_1$, $H_2$ are the two Higgs doublets, $g$ and $g'$ denote the
$SU(2)$ and $U(1)$ gauge coupling constants, respectively, and
$\epsilon_{12}=1$. The terms in~(\ref{higgspot}) relative to the
neutral Higgs fields in $H_1$ and $H_2$
are explicitly reported in~\cite{US}; see note in the Appendix.

The three mass parameters $\mu_1^2,\mu_2^2,\mu_3^2$, at $M_X$ 
given by
\beq
\mu_1^2(M_X) = \mu_2^2(M_X) = m^2+\mu^2, \hspace{1cm}
\mu_3^2(M_X) = -\left( A-1 \right)m \mu\,,
\eeq
are crucial for the breaking of $SU(2)\times U(1)$. The requirement
that the minimization of the potential~(\ref{higgspot}) yields the
correct vacuum provide us with two relations linking the two
vacuum expectations values $v_1,v_2$ (with $v_2$ and $v_1$ giving
rise to the mass of up and down quarks, respectively) to
$\mu_1^2,\mu_2^2,\mu_3^2$ and ultimately to the initial parameters
$\mu,m,M,A$. The two conditions obtained through the minimization
of~(\ref{higgspot}) can therefore be used to trade two of these
initial parameters, as for example $\mu$ and $A$, for
$M_Z$ ($M_Z=1/2 (\ggp)(v_1^2\!+v_2^2)$) and $\tb$ .

Thus, one is left with $\tb$,~$(m,M)$ as independent
parameters of the MSSM, in addition to those already present in
the SM. The choice of $m$ and $M$ as independent parameters and
$\mu$ and $A$ as derived ones is obviously completely arbitrary.
For fixed values of $m_t$ and $\tb$, not all points in the
2-dimensional space $(m,M)$ provide physical realization of the MSSM.

To begin with, the minimization of the potential~(\ref{higgspot})
may not provide an adequate approximation to the minimization of the
full scalar potential. One has to guarantee that no other dangerous
minima breaking charge and/or colour appear below the correct
one. To this aim one can examine each point of the $(m,M)$-space
by means of a set of necessary conditions
among which, only those relative to not too large yukawa couplings can
become sufficient~\cite{GUNHABSHER}.
So far, this is still the best
known procedure to avoid charge- and colour- breaking minima
and the one also used for the present analysis. For all the other
tests which each point of the space $(m,M)$ has to pass before being
accepted as a viable one, the reader is referred to~\cite{US}. When
a point
qualifies as such, $\mu$ and $A$ can be evaluated through the
minimization conditions of~(\ref{higgspot}). Once gauge and yukawa
coupling at $M_X$ are also calculated, the initial conditions for the
RG equations relative to the full low-energy supersymmetric spectrum
are completely specified.

No corrections to the RG-improved tree-level higgs
potential~(\ref{higgspot}) are included in this analysis.
This approximation may not be completely
satisfactory
in some
regions of the supersymmetric parameter space,
in particular in those where some fine-tuning of parameters may be
required. We try, in general, to avoid these regions; see choice of
values for $\tb$, for $m_t=150\,$GeV. For the remaining ones,
we believe that this procedure can
provide
good indications of the complex interplay of the
different contributions to $\bsgamma$. Within this approximation, the
running masses of the
the physical Higgs bosons have the simple structure:
\beq
m_{H_1^0,H_2^0}^2 = {1 \over 2} \left[ m_{H_3^0}^2 + M_Z^2
 \pm \sqrt{(m_{H_3^0}^2 + M_Z^2)^2 - 4m_{H_3^0}^2 M_Z^2 \cos^2 2
 \beta }\ \right] \,,
\label{h12mass}
\eeq
\beq
m_{H^\pm}^2  =  M_W^2 + m_{H_3^0}^2\,,
\label{hpmass}
\eeq
where in turn, $m_{H_3^0}^2$ is given by a combination of the
mass parameters in~(\ref{higgspot}):
\beq
m_{H_3^0}^2   =
   \mu_1^2 + \mu_2^2 \ = \ \frac{2 \mu_3^2}{\sin 2 \beta}\,,
\label{h3mass}
\eeq
with the second equality induced by the requirement of radiative
breaking of $SU(2)\times U(1)$.

The value of gauge and yukawa couplings at $M_X$ are the remaining
ingredient needed in order to integrate the RG equations for the
supersymmetric parameters. If no threshold effects for supersymmetric
particles are considered, as done in the present analysis, gauge and
yukawa couplings obey a set of RG equations not coupled to those for
the supersymmetric parameters. Supersymmetric particles
contribute to the value of their RG equation coefficients, at energies
above $M_{\rm SUSY}$. Below this scale the RG equations for the
2HDM are used. At this same scale, chosen here to be $2 M_Z$, the
evolution down of all the remaining
supersymmetric parameters is also stopped.

Thus, following  this procedure, the evolution of the low-energy
couplings
\beq
\as(M_Z) = 0.114,\hspace*{0.7cm}  \sin^2 \theta_W = 0.233,
                  \hspace*{0.7cm}  \ale(M_Z)= 1/128
\label{input}
\eeq
allows to obtain the grand-unification scale $M_X$ and the common
value of the gauge couplings at $M_X$, $\alpha_{\rm GUT}$. The
chosen value of $\as(M_Z)$ in~(\ref{input}), small when compared
with the LEP results~\cite{ALPHAS}, is well within the currently
allowed values of $\as(M_Z)$; see the world average
$\as(M_Z) = 0.118 \pm 0.007$ reported in ~\cite{ALPHAS}. Changes
in the value of $\as(M_Z)$ may produce (slightly) different values
of $M_X$ and $\alpha_{\rm GUT}$. These changes are, however, almost
unconsequential for the evaluation of the supersymmetric mass
spectrum. Since we are interested in the global features of this
spectrum, obtained over a wide region of the $(m,M)$-space, the
effect of a variation of $\as(M_Z)$ can be compensated by a small
shift in the value of $M$.

Similar is, in principle, the procedure for the evaluation of the
yukawa couplings at $M_X$. After fixing the parameter $m_t(M_Z)$, a
large uncertainty exists on the value of the running mass $m_b(M_Z)$,
which corresponds to the wide range of validity for the physical
bottom quark mass, $4.6 \!- \!5.3\GeV$~\cite{PDB}. This uncertainty
can be exploited
to obtain unification of the yukawa couplings at $M_X$.
%
The usually imposed $SU(5)$-type of unification requires
equality of
bottom and tau couplings,
$h_b(M_X) = h_\tau(M_X)$.
Not clear is how stringent this condition has to be considered.
The consequences of a strict
unification have been recently analyzed~\cite{YUKUNIF}.
Two-loop evolution equations for gauge and yukawa couplings are used
in these calculations as well as an improved approximation
of $M_{\rm SUSY}$, depending on one of the two free supersymmetric
parameters remaining after fixing $m_t$ and $\tb$. Strong
constraints are obtained for the allowed values of $\tb$,
for $m_t$ fixed and the physical bottom mass varying within
the range $4.6-5.3\,$GeV.

We adopt here the more conservative approach of
ref.~\cite{OLECH}. It was shown in this paper that a relaxation of
the yukawa couplings unification at the level of 15-20\%
still allows all possible values of $\tb$. This procedure
yields the same SUSY mass spectrum which one obtains by imposing a more
precise yukawa couplings unification (at the 1\% level), but allowing
the physical bottom mass to go beyond the range of
$4.6 \!-\!5.3\GeV$. As in~\cite{OLECH}, we consider running bottom
masses, at $M_Z$, as high as $3.6\!-\!3.7\GeV$, for $m_t=150\GeV$.

Appropriate lower bounds, corresponding as closely as possible to the
current experimental limits, have to be imposed on the supersymmetric
masses thus calculated. Since the negative searches of supersymmetric
particles rely on assumptions which may not be valid throughout
the full range of the MSSM parameter space, one can only try to
obtain a rough match of the experimental situation.

While it seems realistic to implement the limits at $45\GeV$ imposed
by LEP~I on almost all particles, the CDF limits on squarks and gluino
masses pose some problem~\cite{CDFBOUNDS}. The limits at $126$
and $141\GeV$ respectively, are obtained under the assumptions of a
massless photino, complete degeneracy among squarks and
under the assumption that squarks and gluinos decay directly
into the lightest supersymmetric particle, without intermediate decays
into charginos and neutralinos. When the possibility of these cascade
decays is considered, for one particular choice of the supersymmetric
parameters, it is found that the lower limit on squark
masses, now a function of $m_{\wti g}$, disappears for big enough
values of the gluino mass~\cite{CDFBOUNDS}.  Charginos and
neutralinos can be light enough in the MSSM to give substantial
branching ratios for cascade decays of not too light squarks and
gluinos. Thus, we adopt the strategy of imposing an ``average bound''
of $120\GeV$ and $100\GeV$ for gluino and squarks masses,
respectively.

We make an exception for the lightest eigenvalue of the up-squark mass
matrix, or stop (${\wti u}_1$). Given the presence of the parameter
$m_t$ in the left-right entries of this matrix, stops
can indeed be rather light. The CDF limit does not apply to this case
since the decay mode ${{\wti u}_1} \to t + {{\wti \chi}_1}^0$ is
forbidden or highly disfavoured (depending on the value of $m_t$) and
one can only rely on the limits from LEP. This limit has only recently
been brought up to $39\GeV$~\cite{STOPBOUND}. In view of possible
improvements we already impose a lower limit of $45\GeV$.

The limit on the mass of the lightest eigenvalue of the down-squark
mass matrix, or sbottom (${\wti d}_1$),
deserves also some comments. It is sometimes argued that large values
of $\tb $ in the left-right entries of this matrix can give rise to
sbottom masses as small as, or smaller than, the stop mass.
This point, however, has never been explicitly checked in any previous
supersymmetric search.  We shall report later on, at the end of this
section, on the negative evidence we obtain for
light ${\wti d}_1$, at least up to the value of $\tb$ we
consider. Moreover, since the CDF searches, based on the
decay mode ${{\wti d}_1} \to b + {{\wti \chi}_1}^0$ should be
effective in this case, we impose also for this mass the limit
of $100\GeV$.

Finally, we assume a lower bound of $45\GeV$ on the mass of the
lightest neutral Higgs $m_{H_2^0}$ (for a summary of the lower
limits obtained for this mass by the four LEP Collaboration, see for
example~\cite{HIGGSBOUND}) and a bound of $20\GeV$ on the lightest
neutralino mass (induced in this model by the chargino
limit). No lower bound has to be imposed on the charged Higgs mass
since the theoretical low-energy value for
$m_{H^\pm}$~(\ref{hpmass}), already exceeds the
lower limit coming from LEP.

In short, the lower bounds imposed in the remaining of this paper
are:
\bea
    & {\phantom{{\rm SET I}} }  &           \nn \\[-1.0ex]
                            &     &
m_{\wti g}           \ \     >   120\GeV,  \quad\quad
m_{{\wti d_1},{\wti u_2}}\!\!\!
                     \       >   100\GeV,  \quad\quad
m_{\wti u_1}                 >  \ 45\GeV,  \quad\quad\quad\quad \nn \\
                          &      &
m_{\wti\chi^-_2}  >\ \ 45\GeV,   \quad\quad
m_{\wti\nu_1}\ \  >\,\ 45\GeV,   \quad\quad
m_{\wti l_1}   \  >\   45\GeV,   \quad\quad\quad\quad      \nn  \\
                          &      &
m_{H_2^0}         >\ \ 45\GeV,   \quad\quad
m_{\wti\chi^0_1}
             \, \ >\,\ 20\GeV.   \quad\quad\quad\quad
\label{expbounds}
\eea
\newpage

\subsection{Parameter Space}

In the remaining of this section we show the supersymmetric
parameter space explored for $m_t=150\GeV$ and different values of
$\tb$ in general from $3$~to~$3$ and up to the value
$35$ when searching for $m_{H^\pm} < m_t$. The value $m_t=150\,$GeV
seems representative enough to discuss typical features of the MSSM
and results to be expected
for $\BR(\bsgamma)\vert_\MSSM$. The range of $\tb$ is
also sufficiently wide to allow a survey of possible realizations of
the MSSM obtained in the regime $h_t >> h_b$
up to the one where it is $h_t \gtap h_b$. Smaller values of $\tb$,
as well as larger ones, for which $h_t$ approaches too closely $h_b$,
are avoided. The error due to the fact that no loop corrections to the
tree-level potential~(\ref{higgspot}) are included in this analysis,
may not be negligible in these cases.

Thus, for this chosen value of
$m_t$ and fixed values of $\tb$, the region of the 2-dimensional
space $(m,M)$ within the limits
$0\!<\!m\!<\!500\GeV$, $-250\!<\!M\!<\!250\GeV$ is scanned picking
up points at regular intervals of $6\GeV$ in both directions. Each
point is then tested to verify whether it allows physical realization
of the MSSM and the full low-energy spectrum is calculated if this
turns out to be the case.

We show in fig.~\ref{parspace} the results of this scanning. No
realizations of the MSSM with radiative breaking of $SU(2)\times U(1)$
are obtained in the white areas of this figure,
while the regions where these
realizations are possible are covered by dots. It should be mentioned
here that the points shown in fig.~\ref{parspace} can be
``multiple'' points. Given the non-linearity of the relations among the
original parameters induced by the requirement of radiative breaking of
$SU(2)\!\times\!U(1)$ different solutions for the
pair $(\mu,A)$ are possible when $m$,~$M$,~and~$\tb$ are
fixed. The dark and light dots distinguish regions allowed and
forbidden by the lower limits~(\ref{expbounds}). The observation that
wider regions of parameter space are excluded by the same
bounds~(\ref{expbounds}), for increasing values of $\tb$,
suggests the appearance of lighter supersymmetric masses
when going from $\tb =3$ to 30.

The points of fig.~\ref{parspace}, are plotted in
fig.~\ref{mumtwo} in the plane $(\mu_R,M_2)$, where $\mu_R$ is the
renormalized values of the parameter $\mu$ and $M_2$ the renormalized
value of the $SU(2)$-gaugino mass $M$. Of these two parameters,
entering in the chargino mass matrix, $M_2$ is related to $M$ by a
constant factor (depending on the input~(\ref{input})) $\sim0.8$ and
$\mu_R$ is not too dissimilar from $\mu$. In order to focus on the
smaller values of $\mu_R$, not all the points of fig.~\ref{parspace}
are shown in this figure: the full range $-250<M<250\,$GeV is covered
in the plots of fig.~\ref{mumtwo}, whereas restrictions on the
$m$-range are performed for the different values of $\tb$. When the
full range $0<m<500\,$GeV is considered, values of $\vert \mu_R\vert$
as big as $700\GeV$ are obtained for $\tb=3$ and $500\GeV$ for
the larger values of $\tb$. Superimposed in fig.~\ref{mumtwo}
are the contour lines for $m_{{\wti \chi_2}^-} = 45\GeV$
(dashed lines) and $m_{{\wti \chi_2}^-} = 90\GeV$ (solid lines),
respectively the lower limit on the chargino mass obtained at LEP~I
and the limit which LEP~II may impose.

\begin{figure}[p]
\begin{center}
\epsfxsize=16 cm
\leavevmode
\epsfbox[18 140 534 732]{cons1.ps}
\end{center}
\caption[f5]{\small{The supersymmetric parameter space for
$m_t=150\GeV$ and different values of $\tb$. The white regions
are forbidden in the MSSM, the lighter dots delimited by the solid
lines indicate the regions allowed in the MSSM, but forbidden by the
lower bounds~(\ref{expbounds}), the darker dots the regions where
realistic realizations of the MSSM, compatible with~(\ref{expbounds}),
are obtained.    }}
\label{parspace}
\end{figure}

\begin{figure}[p]
\begin{center}
\epsfxsize=16cm
\leavevmode
\epsfbox[18 140 534 732]{cons2.ps}
\end{center}
\caption[f5]{\small{Points of the parameter space shown in
  fig.~\ref{parspace} plotted in the plane $(\mu_R,M_2)$.
  Light and dark dots indicate possible
  realizations of the MSSM, forbidden or allowed by the lower
  limits~(\ref{expbounds}). Superimposed are the contour lines
  $m_{{\wti \chi_2}^-} = 45\GeV$ (dashed lines) and
  $m_{{\wti \chi_2}^-} = 90\GeV$ (solid lines).       }}
\label{mumtwo}
\end{figure}

The strong correlations introduced by the radiative breaking of
$SU(2) \times U(1)$ are visible in this figure. For
$\tb=9$-$30$ and the smallest values of $m$ considered,
the solution $\mu \gtap M$ is obtained. For increasing values of
$m$, $\vert\mu\vert$ increases for $M<0$ and decreases for $M>0$,
creating the diagonal bands of fig.~\ref{mumtwo}.

We observe that
for $\tb \gtap 9$ these bounds are wider for $M<0$ than for
$M>0$. Altogether different are the solutions obtained for $\tb=3$:
very narrow are the bands one obtains for $\mu_R>0$, whereas large
distributions of points are present for negative $\mu_R$. For
$\tb \gtap 9$ a different type of solutions is also present: small
values of $\mu$ are obtained even for relatively large values of $M$
and not too large values of $m$. These solutions, tend to align along
smaller and smaller values of $\mu$, for increasing $\tb$.

The differences observed for the solutions obtained below and above
$\tb = 9$ affect the distributions of masses obtained; see discussion
on charged Higgs and chargino masses.

In the following subsections we shall show the typical ranges of
masses obtained for the supersymmetric particles relevant to the
decay $\bsgamma$ and the location of some interesting intervals of
these masses in the $(m,M)$-plane.

\subsection{Charged Higgs Mass}

We start here with the charged Higgs. We show in fig.~\ref{cnhiggs}
the position occupied in the $(m,M)$-plane by intervals of
progressively heavier ${H^\pm}$. Masses
below $100\GeV$ are indicated by the vertical lines closer to the
forbidden regions, whereas successive intervals of masses,
$100\!-\!150\GeV$, $150\!-\!200\GeV$, $200\!-\!250\GeV$ and
$250\!-\!300\GeV$, form layers, distinguished in fig.~\ref{cnhiggs}
by different types of lines, wrapped one on top of the other around
the innermost one. By adding
additional layers with larger values of masses, all the points in
fig.~\ref{parspace} would be recovered.

The regular structure obtained for the $m_{H^\pm}$ contour levels
can be understood upon inspection of the decomposition of
$m_{H^\pm}^2$ in terms of the four initial parameters:
\beq
m_{H^\pm}^2=
   c_1 \mu^2 + c_2 M^2 +  c_3 m^2 + c_4 (Am)^2 + c_5 AMm \,.
\label{higgsdecomp}
\eeq

\begin{figure}[p]
\begin{center}
\epsfxsize=16cm
\leavevmode
\epsfbox[18 140 534 732]{cnch.ps}
\end{center}
\caption[f5]{\small{Regions where different intervals of
$m_{H^\pm}$ are located. The slanted lines closest to the regions
forbidden in the MSSM indicate
the intervals of masses between $M_W$ and~$100\GeV$. The following
circular regions correspond to the intervals $100\!-\!150\GeV$,
$150\!-\!200\GeV$, $200\!-\!250\GeV$ and $250\!-\!300\GeV$,
respectively.
No lower bounds on supersymmetric masses
are imposed on this figure, but the contours of the regions
removed by~(\ref{expbounds}) (solid lines) are
drawn as a reminder.      }}
\label{cnhiggs}
\end{figure}

\begin{figure}[tb]
\begin{center}
\epsfxsize=12.4cm
\leavevmode
\epsfbox[60 360 523 751]{thb150_cnst.ps}
\end{center}
\caption[f5]{\small{Allowed ranges of masses for the charged Higgs
$H^+$ into which the top quark can decay in the MSSM ($t\to bH^+$). The
obtained values of $m_{H^+}$ are shown for different values of
$\tb$, from 3 to 35. The lower bounds~(\ref{expbounds}) are
imposed in this search.                            }}
\label{cnthb}
\end{figure}

The value of the coefficients $c_i$, reported explicitly for
$\mu_1^2-m^2$, $\mu_2^2-m^2$
($m_{H^\pm}^2= \mu_1^2+\mu_2^2 +M_W^2$)
in the following table,

\vspace*{0.3cm}
\hspace{0.8cm}
\mbox{
\vbox{\offinterlineskip
\halign{
\strut#
        &     \quad #  &
        \quad \hfil #  &
        \quad \hfil #  &
        \quad \hfil #  &
        \quad \hfil #  &
        \quad \hfil #  &
        \quad \quad #  \cr
\noalign{\hrule \vskip 2pt }
\omit & \hfil$\mu^2$\hfil &\hfil$M^2$\hfil
      & \hfil$m^2$\hfil   &\hfil$(Am)^2$\hfil &\hfil$AMm$\hfil & \cr
\noalign{\vskip 2pt \hrule \vskip 2pt \hrule \vskip 2pt}
%
$\mu_1^2-m^2$ &$1.142$&$ 0.488$&$0.994$&$-0.002$&$0.006$&$\tb=3$ \cr
$\mu_2^2-m^2$ &$1.142$&$-1.946$&$0.049$&$-0.116$&$0.451$&        \cr
\noalign{\vskip 2pt \hrule \vskip 2pt}
%
$\mu_1^2-m^2$ &$1.208$&$ 0.370$&$0.950$&$-0.015$&$0.049$&$\tb=9$ \cr
$\mu_2^2-m^2$ &$1.208$&$-1.778$&$0.136$&$-0.121$&$0.468$&        \cr
\noalign{\vskip 2pt \hrule \vskip 2pt}
%
$\mu_1^2-m^2$ &$1.172$&$ 0.139$&$0.863$&$-0.038$&$0.126$&$\tb=15$\cr
$\mu_2^2-m^2$ &$1.172$&$-1.756$&$0.143$&$-0.119$&$0.460$&        \cr
\noalign{\vskip 2pt \hrule \vskip 2pt}
%
$\mu_1^2-m^2$ &$1.122$&$-0.114$&$0.762$&$-0.062$&$0.203$&$\tb=20$\cr
$\mu_2^2-m^2$ &$1.122$&$-1.744$&$0.145$&$-0.116$&$0.451$&        \cr
\noalign{\vskip 2pt \hrule \vskip 2pt}
%
$\mu_1^2-m^2$ &$1.058$&$-0.411$&$0.635$&$-0.085$&$0.278$&$\tb=25$\cr
$\mu_2^2-m^2$ &$1.058$&$-1.731$&$0.145$&$-0.113$&$0.437$&        \cr
\noalign{\vskip 2pt \hrule \vskip 2pt}
%
$\mu_1^2-m^2$ &$0.977$&$-0.738$&$0.484$&$-0.105$&$0.340$&$\tb=30$\cr
$\mu_2^2-m^2$ &$0.977$&$-1.715$&$0.145$&$-0.109$&$0.422$&        \cr
\noalign{\vskip 2pt \hrule \vskip 2pt}
%
}}
}

\noindent
and the requirement $\vert \mu\vert \gtap \vert M \vert$, imposed by
the radiative breaking of $SU(2)\times U(1)$ (see
fig.~\ref{mumtwo}), explains the almost semi-circular shapes
obtained for the $m_{H^\pm}$ contour levels.

The steady decrease of the coefficients for $m^2$ and the overall
decrease of the weight of the remaining contributions, for
increasing $\tb$, explains the other interesting feature shown in
fig.~\ref{cnhiggs}: for the larger values of $\tb$,
the low ranges of masses invade almost completely the $(m,M)$-region
considered here, while for $\tb =3$, the same region can be
covered only when going to masses $m_{H^\pm}$ as big
as $850\!-\!900\GeV$.

The first two layers in fig.~\ref{cnhiggs} represent roughly the
allowed phase space for the decay $t\to bH^+$, for $m_t=150\GeV$. The
interplay between the expansion of the mass intervals in
fig.~\ref{cnhiggs} and the differences in shape and size of the regions
removed by the bounds~(\ref{expbounds}) for increasing $\tb$,
makes the decay $t\to bH^+$ less ``probable'' not in the case of
$\tb=3$, where less points are found in the first two layers,
but for some intermediate value of $\tb$ between 3 and 15.

We plot in fig.~\ref{cnthb} this phase space in the plane
$(m_{H^+},\tan\beta)$, for specific choices of $\tb$, from
$3$~to~$35$. The values of $m_b(M_Z)$ considered here range from
$3.5$ to $3.7$. It will be shown in the next section, in
fig.~\ref{rsthb}, how this phase space is reduced by the requirement
that $\BR(\bsgamma)\vert_\MSSM$ be limited to the range of values
compatible with the SM prediction. Since the region of parameter space
probed here is relatively small, we use in this as well as in
fig.~\ref{rsthb} a finer grid with a spacing of only $2\GeV$ while
scanning the same portion of the $(m,M)$-plane. The increased number
of points obtained and the hidden degree of freedom remaining when
$\tb$ and $m_{H^+}$ are fixed, make each line obtained look
almost like a continuous one. The isolated points observed at the
beginning of some of these lines are due to the appearance of isolated
corners at the edge of each layer when the lower
limits~(\ref{expbounds}) are imposed.

For this set of lower bounds, the value of $\tb$ where the decay
$t\to bH^+$ is less probable seems indeed to be $\tb \sim 9$:
starting from $\tb =3$, the allowed range of $m_{H^\pm}$ decreases,
reaching a minimum for $\tb = 9$, and increases then again for
larger values of $\tb$. The results described here differ from those
presented in~\cite{OLECH}, where rather heavy charged Higgses are
obtained except when $h_b \sim h_t$. This disagreement may be due to
the different assumptions in the two calculations, although the
precise reason has still to be pinned down.

\subsection{Chargino and Lightest Up-Squark Masses}

We consider here how different intervals of masses for the lightest
chargino ${\wti \chi_2}^-$ are distributed in the $(m,M)$-plane. The
results are shown in fig.~\ref{cncharg}. For clarity, only
two ranges, $m_{{\wti \chi_2}^-}\!<\!50\GeV$ and
$50\!\leq \!m_{{\wti \chi_2}^-}\!\leq\!100\GeV$ are shown in this
figure. The range with the smaller values of masses occupies the
central part of each plane, while the second interval is split almost
symmetrically around the previous one. The full $(m,M)$-region
considered here is covered with equally symmetric strips of
increasingly larger values of $m_{{\wti \chi_2}^-}$.

%
The modest dependence on $m$ of the mass intervals shown in
fig.~\ref{cncharg} can be explained with the correlation
$\vert \mu\ \vert \sim \vert M \vert $ observed for small values of
$m$ and with the increased probability of having ${\wti \chi_2}^-$ as
a state of pure $W$-ino for increasing $m$ (when $\vert \mu \vert$
can substantially exceed $\vert M\vert$).
Enlargement of the almost horizontal strips shown in
fig.~\ref{cncharg} for large values of $m$ are obtained for
$\tb =9-30$, but are not present for $\tb =3$.

The patterns in the correlations between $\mu_R$ and $M_2$ observed
in fig.~\ref{mumtwo} should also explain the striking similarity
in the results obtained for $\tb =9-30$ (except for a change
in size of the region removed by~(\ref{expbounds}) in the case
$\tb=30$) and the differences between
these results and those obtained for $\tb <9$.

\begin{figure}[p]
\begin{center}
\epsfxsize=16cm
\leavevmode
\epsfbox[18 140 534 732]{cnchi.ps}
\end{center}
\caption[f5]{\small{Regions of the parameter space $(m,M)$ where
 $m_{{\wti \chi_2}^-}\!<\! 50\GeV$ (central strips with horizontal
 lines) and where $50\!\leq\! m_{{\wti \chi_2}^-}\!\leq\! 100\GeV$
 (the two almost symmetrical strips with vertical lines). The lower
 limits~(\ref{expbounds}), when imposed, remove the regions
 enclosed by the solid lines.                   }}
\label{cncharg}
\end{figure}

\begin{figure}[p]
\begin{center}
\epsfxsize=16cm
\leavevmode
\epsfbox[18 140 534 732]{cnst.ps}
\end{center}
\caption[f5]{\small{Same as in figs.~\ref{cnhiggs}~and~\ref{cncharg} for
$m_{{\wti u_1}}$. The internal regions with slanted lines, present only
for $\tb\!=\!3,25$~and~$30$, indicate the intervals
$m_{{\wti u_1}}\!\leq\! 50\GeV$. The remaining two types of regions, with
horizontal and vertical lines correspond to the intervals
$50\!\leq \!m_{{\wti u_1}}\!\leq \!100\GeV$ and
$100\!\leq \!m_{{\wti u_1}}\!\leq \!150\GeV$, respectively.
Outlined are also the areas removed
by~(\ref{expbounds}).   }}
\label{cnstop}
\end{figure}

We observe also that the constraint
$m_{{\wti \chi_2}^-}> 45\GeV$ is, for not too small $m$, the most
severe among the lower bounds~(\ref{expbounds}), for $\tb = 9-30$.
In these cases, the shape of the region removed by~(\ref{expbounds})
corresponds practically to the contour line
$m_{{\wti \chi_2}^-} = 45\GeV$. In contrast, in the case
$\tb=3$, for not too small $m$, this shape is given by the
gluino bound $m_{\wti g}>120\GeV$. For all values of $\tb$,
sizable regions of parameter space with
$45 <m_{{\wti\chi_2}^-}<50\GeV$ remain after imposing the lower
limits~(\ref{expbounds}). They are also present
at large $m$, where $m_{H^-}$ can range from $850\!-\!900\GeV$,
for $\tb=3$, to $200\!-\!250\GeV$, for $\tb=30$.

It should therefore become clear what was observed in the introduction:
the chargino contributions to $\bsgamma$ can definitely exceed the
Higgs contributions for large values of $m$ and small values of
$\tb$. Chargino contributions exceeding the SM and Higgs
contributions can in principle be expected also for larger values
of $\tb $, provided not too heavy stop squarks are obtained in the
same regions.

The distribution of different ranges of mass for a first and second
generation of squarks is expected to follow very regular patterns. The
expected shapes should be ellipsoidal with a smaller vertical axis
(compare with those obtained for $m_{H^\pm}$) due to the large gluino
contribution to the renormalization of squarks masses. The
coefficients for the decomposition of the squared masses of first and
second generation of squarks in term of the original four
supersymmetric parameters (for the choice~(\ref{input}))
are explicitly given in ref.~\cite{IO}.
Deviations from these simple patterns are expected for the lightest
up- and down- mass eigenvalues, due to the presence of potentially
large left-right mixings in the squarks mass matrices.

The result of the investigation of the lightest up-squark mass matrix
is shown in fig.~\ref{cnstop}. Only the intervals of masses
$0\!-\!50\GeV$, $50\!-\!100\GeV$ and $100\!-\!150\GeV$ are plotted
here since the figures tend to become rather complex if intervals with
larger values of masses are also included. The values of $m_{\wti u_1}$
can indeed be rather small and for the two largest values of
$\tb$ considered in this analysis masses $m_{{\wti u}_1}$ as
small as $45\!-\!50\GeV$ appear in coincidence with equally small
values of $m_{{\wti \chi_2}^-}$.

\subsection{Gluino, Neutralino and Lightest Down-Squark Masses}

We come now to a brief analysis of the masses relevant to the
gluino and neutralino contribution to $\bsgamma$.

The values of $m_{\wti g}$ we are dealing with in this study are
easily obtained through a scaling of $M$ by a factor $\sim 2.6$
(again dependent on the choice~(\ref{input})).

Patterns of masses absolutely similar to those shown in
fig.~\ref{cncharg} for the mass of the lightest chargino
${\wti \chi_2}^-$ are obtained for the lightest neutralino
${\wti \chi_1}^0$, provided the value of each interval of mass is
divided by a factor $\sim 2 $.

Gluinos and neutralinos are coupled to down-squarks ${\wti d}_i$ when
exchanged as virtual particles inside the loop mediating the decay
$\bsgamma$. Strong deviations from the
simple distributions of masses obtained for the first and second
generation of squarks are expected for the sbottom,
when large values of $\tb$ are considered.

\begin{figure}[p]
\begin{center}
\epsfxsize=16cm
\leavevmode
\epsfbox[18 140 534 732]{cnsb.ps}
\end{center}
\caption[f5]{\small{Same as in
figs.~\ref{cnhiggs},~\ref{cncharg}~and~\ref{cnstop} for
$m_{{\wti d_1}}$. The regions with slanted, vertical, slanted and
horizontal lines, indicate the intervals
$50 \!\leq\! m_{{\wti d_1}}\!\leq\! 100\GeV$,
$100\!\leq\! m_{{\wti d_1}}\!\leq\! 150\GeV$ and
$150\!\leq\! m_{{\wti d_1}}\!\leq\! 200\GeV$, respectively.
No solutions are obtained for $ m_{{\wti d_1}}\!\leq\!50\GeV$ and in
the interval $50\!-\!100\GeV$
all the points cluster around the largest value
of $m_{{\wti d_1}}$, for all values of $\tb$ shown. Superimposed
are the contours of the regions removed by~(\ref{expbounds}).   }}
\label{cnsbott}
\end{figure}

We show in fig.~\ref{cnsbott} some of the allowed values of masses
obtained for the sbottom squark. We search for the intervals
$ 0 \leq m_{{\wti d_1}}\leq 50 \GeV$,
$50 \leq m_{{\wti d_1}}\leq 100\GeV$,
$100\leq m_{{\wti d_1}}\leq 150\GeV$ and
$150\leq m_{{\wti d_1}}\leq 200\GeV$.
No small values of this mass,
i.e. $0\ltap m_{{\wti d_1}}\ltap 50\GeV$, are obtained, even for the
largest value of $\tb$ considered. Moreover, also in the first
interval of masses, $50\!-\!100\GeV$, we obtain values of
$m_{\wti d_1}$ clustering quite closely around $\sim 100\GeV$. These
results justify a posteriori the lower bound on
$m_{{\wti d}_1}$, imposed in our analysis, independently of
$\tb$.

Although the shapes obtained are more regular than those
observed for $m_{\wti u_1}$, the presence of the large left-right
mixing terms in the down-squark mass matrix, becomes visible for large
values of $\tb$. Again, a regular expansion of the mass-intervals
considered, is observed for increasing values of $\tb$.

More variables than the sheer values of the lightest masses, shown in
this section, are responsible for the prediction of
$\BR(\bsgamma)\vert_\MSSM$. Very important is, for example, the
size of the off-diagonal terms responsible for the couplings
${ \wti g}$-${\wti d}_i$-${\wti d}_j$ and
${{ \wti \chi_k}^-}$-${\wti d}_i$-${\wti d}_j$ and the degeneracy
among the different mass eigenvalues entering in the
calculation. The possibility of enhancements/suppressions, which one
may guess by simply analyzing the mass spectrum, has therefore to be
verified for each point of the supersymmetric parameter space considered.
\newpage

\section{The Branching Ratio}

The decay $\bsgamma$, like all $b \to s $ transitions, is characterized
by the presence of two different scales
$m_{\,l}^2 $
and
$m_{\,h}^2$,
with
$m_{\,l}^2 << m_{\,h}^2$.
Within the SM,
$m_{\,h}^2$
indicates generically the mass of $W$ and $Z$
bosons and of the top quark exchanged as virtual particles, while
$m_{\,l}^2 $
is the scale of the remaining light quarks. Large logarithmic terms
$\log\left(m_{\,h}^2/m_{\,l}^2\right)$
appear when QCD corrections to this decay are included.

The customary procedure used to properly resums these large logarithms
is to rely on the use of the effective Hamiltonian
\beq
H_{eff}(b \to s)=  - \sum_{j} \
       C_{j}\left(\frac{\mu^2}{m_{\,h}^2}\right) \,
       O_{j}\left(\frac{\mu^2}{m_{\,l}^2}\right) \,,
\label{effham}
\eeq
obtained from the Hamiltonian of the underlying theory through an
expansion in inverse powers of
$m_{\,h}^2$.
Except for the link coming from the common scale $\mu$, the
dependence on the heavy degrees of freedom is formally separated
from the dynamics of the light fields to which the theory is now
reduced. This dependence is
recovered through insertions on the light fields of the operators
in (\ref{effham}) with generalized charges, or Wilson coefficients,
$C_j(\mu^2/m_{\,h}^2)$.
In the following, we indicate operators
as $O_{j}(\mu)$,
and coefficients as $C_j(\mu)\vert_{\rm SM}$, with a subscript
as a reminder of the underlying theory. The normalization
of the operators $ O_{j}(\mu)$ is such that the coefficients
$C_j(\mu)\vert_{\rm SM}$ match at $\mu =M_W$ the matrix elements
obtained in the SM at the zeroth order in
$\as$. The value of these coefficients at any scale $\mu$ is
then obtained through
RG techniques: the evolution down from
$M_W$ takes appropriate care of switching on the SU(3) interaction
and correctly resums the large logarithms
$\log{\mu^2/m_{\,h}^2}$.

At a fixed order in the perturbative expansion, the amplitudes for
these $b \to s$ transitions depend on the renormalization scale $\mu$.
Although one may want $\mu \sim {\cal O}(m_b)$, the precise value of
this scale
is unknown. The uncertainty due to this arbitrariness
is the biggest one in the theoretical
determination of $\BR(\bsgamma)\vert_{\rm SM}$ at the LO in QCD and
could be reduced if higher order QCD corrections to all elements
in (\ref{effham}) were included.

Up to eight operators have been found to contribute to the decay
$\bsgamma$. Nevertheless,
the basis has often been truncated to subsets of three operators:
\bea
O_{1}  &=&  \left(\bar{c}_{L \beta} \go{\mu} b_{L \alpha}\right)
           \left(\bar{s}_{L \alpha} \gu{\mu} c_{L \beta}\right)
                                                        \nn \\[1.15ex]
O_{2}  &=&  \left(\bar{c}_{L \alpha} \go{\mu} b_{L \alpha}\right)
           \left(\bar{s}_{L \beta} \gu{\mu} c_{L \beta}\right)
                                                            \\
O_{7}  &=&  \frac{e}{16\, \pi^2} \, m_b
           \, \left( \bar{s}_\alpha \, \sigma^{\mu \nu}
           \, {\rm P_R^{\phantom{A}}}  b_\alpha \right)\, F_{\mu \nu}\,,
                                                        \nn
\label{basis}
\eea
or alternatively $O_2$, $O_7$ and $O_8$, with
\beq
O_{8} =  \frac{\gs}{16\, \pi^2} \, m_b
           \, \left( \bar{s}_\alpha \, \sigma^{\mu \nu}
           \, {\rm P_R} \, T^A_{\alpha \beta}
           \, b_\beta \right) \, G^{A}_{\mu \nu}\, .
\label{gluonoper}
\eeq
Relatively simple and closed forms for the coefficient
$C_7(\mu)\vert_{\rm SM}$ are
then obtained as function of the initial values
$C_1(M_W)\vert_{\rm SM}$:
\beq
C_7 (\mu)\vert_{\rm SM} = \eta^{-\frac{16}{23}}
                           \left\{ C_7(M_W)\vert_{\rm SM}
                          -  \left[
       \frac{58}{135} \left( \eta^{\frac{10}{23}} \!- 1 \right) \, +
       \frac{29}{189} \left( \eta^{\frac{28}{23}} \!- 1 \right)  \,
                             \right]
                          \, C_2(M_W)\vert_{\rm MS} \right\}\, ,
\label{corr1}
\eeq
in the first case
, or
$$
C_7 (\mu)\vert_{\rm SM} = \eta^{-\frac{16}{23}}
                           \left\{ C_7(M_W)\vert_{\rm SM}
        + \frac{8}{3} \left(
              \eta^{\frac{2}{23}} \! - 1 \right) C_8(M_W)\vert_{\rm SM}
     - \, \frac{232}{513} \left(
              \eta^{\frac{19}{23}} \! -1 \right) C_2(M_W)\vert_{\rm SM}
                                \right\}                \nn
$$
\beq
\label{corr2}
\eeq
in the second one. The symbol $\eta$ indicates $\as(\mu)/\as(M_{W})$ and
$C_1(M_W)\vert_{\rm SM}$, $C_2(M_W)\vert_{\rm SM}$,
$C_7(M_{W})\vert_{\rm SM}$ and $C_8(M_W)\vert_{\rm SM}$ are
respectively:
\bea
 C_1(M_W)\vert_{\rm SM} & = &\ \ 0                     \, , \nn \\[1.2ex]
C_2(M_W)\vert_{\rm SM}  & = & \frac{4 G_{F}}
                                {\sqrt{2}} \, K_{cs}^*K_{cb}\, , \nn \\
C_7(M_W)\vert_{\rm SM}  & = & \! -\frac{1}{2} \,
         {\cal A}^\gamma_{\rm SM} \, / \, \frac{e}{16\pi^2} \, , \nn \\
C_8(M_W)\vert_{\rm SM}  & = & \! -\frac{1}{2} \,
         {\cal A}^g_{\rm SM}   \, / \, \frac{\gs}{16\pi^2}  \, ,
\eea
with $K_{ij}$ elements of the Cabibbo-Kobayashi-Maskawa matrix and
${\cal A}^\gamma_{\rm SM}$ and ${\cal A}^g_{\rm SM}$
given in eqs.~40-44 of ref.~\cite{US}.
The branching ratio, depending on $\mu$,
is then obtained as
\beq
 \BR(\bsgamma;\mu)\vert_{\rm SM} =
      \frac{\Gamma(\bsgamma;\mu)\vert_{\rm SM}}
           {\Gamma(b \to c e \nu;\mu)\vert_{\rm SM}}
                \, \BR(b \to c e \nu)\vert_{\rm exp}
\label{branch}
\eeq
where the width $\Gamma(\bsgamma;\mu)\vert_{\rm SM}$ reads:
\beq
 \Gamma(\bsgamma;\mu)\vert_{\rm SM}                      \ = \
      \frac{\alpha \, m_b^5} {\left(4 \pi \right)^4}\,
      \left[ \, C_7(\mu)\vert_{\rm SM} \right] ^2\,. 
\label{width}
\eeq

Estimates of the size of the uncertainty introduced in the
theoretical evaluation of $\BR(\bsgamma)\vert_{\rm SM}$ due to
different truncations of the basis $\{O_j(\mu)\}$ and of the error
with respect to the calculation where the complete basis has been
retained, give values never exceeding 10-15\%~\cite{GRSPWI,MISIAK}.
By roughly the same value is dominated the error due to the
assumption that only two different scales enter in this calculation,
whereas the situation
$m_{\,l}^2 \!<< \!M_W^2 \!< \!m_t^2$
should be considered~\cite{GRICHO}.
The errors introduced by these two types of approximations are well
within the uncertainty due to the dependence of
$\BR(\bsgamma)\vert_{\rm SM}$ on the
renormalization scale $\mu$. Fixing, for example, $m_t=140\GeV$ and
assuming the SM relation
$\vert K_{\rm cb}\vert = \vert K_{\rm ts}\vert$, ratios
$\BR(\bsgamma;\mu\!=\!2.5\GeV) / \BR(\bsgamma;\mu\!=\!10\GeV) $
as big as $1.7$~\cite{ALIGR} are found%
{}~\footnote{The calculation in~\cite{ALIGR} differs from ours since
 includes in the evaluation of $\BR(\bsgamma)\vert_{\rm SM}$ also the
 QCD breemsstrahlung process $\bsgamma g$. Nevertheless, our results
 can be easily compared with those in~\cite{ALIGR} since this additional
 process accounts for an extra factor $K(\mu)$ on the left-hand side
 of~(\ref{width}). The $\mu$ dependence of $K(\mu)$ is completely
 negligible and therefore one should be able to compare the uncertainty
 related to the scale dependence which we shall claim later on, with the
 one obtained in~\cite{ALIGR}.  }.

If to this uncertainty one also adds the one induced by the
unknown value of $m_t$ and the uncertainty in the experimentally
allowed value of $\vert K_{\rm ts}\vert$, it is easy to conclude
that the value of $\BR(\bsgamma)\vert_{\rm SM}$ is rather poorly
known.
The uncertainty of calculations of this branching ratio in extensions of
the SM is, in general, expected to increase, matching at best the one
for the SM prediction.

No new operators contributing to $\bsgamma$ are found in the MSSM,
nor in the 2HDM and no modification due to the presence of extra
particles are introduced in the evolution relations~(\ref{corr1})
and~(\ref{corr2}). Thus, the obvious change with respect to the SM
calculation comes, in both models, in the evaluation of the Wilson
coefficients. In particular,
in the 2HDM the coefficient $C_7(M_W)$ reads:
\beq
C_7(M_W)\vert_{\rm 2HDM}  =  -\frac{1}{2} \, \left[
  {\cal A}^\gamma_{\rm SM} + {\cal A}^\gamma_{H^-}
                                           \right]
                            \, / \, \frac{e}{16\pi^2}\,,
\label{coeffhigg}
\eeq
whereas, in the MSSM, it is:
\beq
C_7(M_W)\vert_\MSSM  =  -\frac{1}{2} \, \left[
  {\cal A}^\gamma_{\rm SM} + {\cal A}^\gamma_{H^-} +
  {\cal A}^\gamma_{\wti{\chi}^-} + {\cal A}^\gamma_{\wti g} +
  {\cal A}^\gamma_{\wti{\chi}^o}           \right]
                            \, / \, \frac{e}{16\pi^2}\, ,
\label{coeffsusy}
\eeq
where ${\cal A}^\gamma_{\rm SM}$,\,${\cal A}^\gamma_{H^-}$
 \,${\cal A}^\gamma_{\wti{\chi}^-}$,\,${\cal A}^\gamma_{\wti g}$
and ${\cal A}^\gamma_{\wti{\chi}^o}$, listed in
eqs.~(40)-(45) of ref.~\cite{US}, indicate the five different
contributions mediating $\bsgamma$ in this model. Thus, the
uncertainty related to the truncation of the basis $\{O_j(\mu)\}$ and
to the renormalization scale dependence 
is, for both models, the
uncertainty of the SM calculation.

Worse is, in contrast, the approximations to two scales,
$m_{\,l}^2 $, $m_{\,h}^2$, with
$m_{\,l}^2 << m_{\,h}^2$,
As shown in the previous section, a rather rich spectrum of masses is
obtained in the MSSM and a wide spread of scales can appear in the
calculation of $C_7(M_W)\vert_\MSSM$. If
one indicates by $m_i^2$ and $m_j^2$ the squared masses of two generic
particles exchanged 
in the 1-loop diagrams, different
possibilities such as
a) $ m_{\,l}^2 \!<< M_W^2 \! < m_i^2 \! < m_j^2 $,
b) $ m_{\,l}^2 \!<< m_i^2 \! < M_W^2 \! < m_j^2 $,
c) $ m_{\,l}^2 \!<< m_i^2 \! < m_j^2 \! < M_W^2 $
may occur. The possibility a) is shared by
the 2HDM, when $m_{H^\pm} \gtap M_W$ and $m_{H^\pm} \neq m_t$, while
the remaining ones are typical of the MSSM (lightest chargino, stop
and lightest neutralino can be well below $M_W$).

Thus, the approximation to two scales acceptable in the SM, and
(possibly) in the 2HDM, may not be equally good in some portions
of the 3-dimensional parameter space of the MSSM. We shall make
nevertheless this approximation. As in
ref.~\cite{US}, we shall neglect renormalization effects for the
neutralino contribution, see next section, since neutralinos can
indeed be rather light. Even this strategy, however, may not be
fully adequate, since down squarks, to which neutralinos are coupled,
are quite heavy.  Nevertheless, the procedure chosen to treat this
contribution, given its smallness, is not likely to represent the
main source of uncertainty related to the identification of all scales
to $M_W$, which comes, presumably from the chargino contribution.

Finally, we shall calculate the coefficient $C_7(M_W)$
``relatively'' to the SM, i.e. choosing a particular value for $m_t$
and imposing the SM relations
$\vert K_{cb}\vert = \vert K_{ts}\vert  $ and
$\vert K_{tb}\vert = \vert K_{cs}\vert = 1$ while scanning the
3-dimensional parameter space $(\tb,m,M)$,
and we shall rely
on the form of the coefficient $C_7(\mu)\vert_\MSSM$ as given
in (\ref{corr1}),
evaluated at an average value of $\mu$, $\mu\sim 5\GeV$.
A quantitative check of the differences one may obtain by using
(\ref{corr2}) will also be made.
The supersymmetric spectrum of masses and couplings, obtained as
described in the previous section, is used for the calculation
of $C_7(M_W)$.

Moreover, assuming that the precision of the MSSM calculation is the
same as the SM one, we shall exclude the regions of this 3-dimensional
parameter space which give
$\BR(\bsgamma)\vert_\MSSM$ outside the range of uncertainty of the
SM result, for the same value of $m_t$ and of the CKM elements.
Since the dependence on the unknown parameter $\mu$ is the most severe
of all uncertainties of the SM calculation, we shall choose the
range obtained when varying the scale $\mu$ from $2.5$ to $10\GeV$,
as range of the allowed values of $\BR(\bsgamma)\vert_\MSSM$.
Thus, for $m_t=150\GeV$, we shall retain only the regions of
parameter space where it
is~\footnote{This interval of allowed values of
 $\BR(\bsgamma)\vert_{\rm SM}$, roughly corresponds to the one
 obtained in ref.~\cite{ALIGR}, when the factor $K(\mu)$ is
 removed.}:
\beq
  2.9\times 10^{-4} <
BR\left(\bsgamma; \mu \!= \!5\GeV \right)\vert_\MSSM
                   < 4.8\times 10^{-4}\,.
\label{allowed}
\eeq
Shifts of about $10\%$ in the position of this
interval can be obtained when values of $\as(M_W)$ different from the
one which can be obtained from choice~(\ref{input}) are used
($\as(m_b)$ is fixed). The choice made here on one side maximizes the
effects of QCD corrections to the $\bsgamma$ decay, on the other,
allows to reach down to lower values of the supersymmetric masses.
For the final value of $\BR(\bsgamma)$, the
two effects tend to compensate. We do not observe significant changes
in the resulting exclusion patterns.

In view of the inevitable comparison between the exclusion patterns
obtained in the 2HDM and the MSSM, which will arise in the next section,
we observe here that, for the same values of $m_t$, $m_{H^\pm}$ and
$\tb$, (\ref{coeffsusy}) and (\ref{coeffhigg}) are related by:
\beq
C_7(M_W)\vert_\MSSM\ \equiv \
      C_7(M_W)\vert_{\rm 2HDM} + \Delta C_7^\prime(M_W)
\eeq
with obvious definition of the symbol $\Delta C_7^\prime(M_W)$. Similar
expressions can be written for the coefficients
$C_8(M_W)\vert_{\rm 2DHM}$ and $C_8(M_W)\vert_\MSSM$. Widths
and branching ratios in the two models are related according to:
\bea
\Gamma(\bsgamma)\vert_\MSSM &  = &
   \phantom{B}
   \Gamma(\bsgamma)\vert_{\rm 2HDM} \
                + \ \Delta \Gamma^{\,\prime} \nn \\
 \BR(\bsgamma)\vert_\MSSM  & = &
   \BR(\bsgamma)\vert_{\rm 2HDM} \
                + \ \Delta B\!R^{\,\prime}
\label{brtwohiggs}
\eea
with $\Delta \Gamma^{\,\prime}$ 
given by:
\beq
\Delta \Gamma^\prime \ \equiv\
      \frac{\alpha \, m_b^5}{\left(4 \pi \right)^4}\,
      \left\{ \left[ \,\Delta C_7^\prime(M_W) \,\right]^2
           \,  +\, 2
              \left[ \,\Delta C_7^\prime(M_W) \,\right]
                       C_7(M_W)\vert_{\rm 2HDM}
   \right\} 
\eeq
and $\Delta B\!R^{\,\prime}$ 
obtained from
$\Delta \Gamma^{\,\prime}$ as in~(\ref{branch}).
The claims made in refs.~\cite{HEWETT,BARGER} correspond to the
limit $\Delta C_7^\prime(M_W) \sim 0 $, i.e. they apply, more
pertinently, to the 2HDM.
\newpage

\section{Results}
%
\subsection{Amplitude and Branching Ratio}

We present in this section a discussion of the numerical results
for $\BR(\bsgamma)\vert_\MSSM$ obtained in this analysis and of
the consequences that a possible measurement of the branching ratio
for this decay compatible with the SM prediction can have for
searches of supersymmetric particles. As already done in the previous
section, we shall limit ourselves to one specific
value of the top quark mass, $m_t=150\GeV$ since
completely similar
features are obtained for different values of $m_t$.

\begin{figure}[p]
\begin{center}
\epsfxsize=16cm
\leavevmode
\epsfbox[18 140 534 732]{phgra.eps}
\end{center}
\caption[f5]{\small{Values of $\BR(\bsgamma)\vert_\MSSM$ obtained
 in the allowed points of the $(m,M)$-plane, compatible with the lower
 bounds~(\ref{expbounds}) (dark dots of fig.~\ref{parspace}),
 for $m_t=150\GeV$ and
 different values of $\tb$. The solid line correspond to
 $\BR(\bsgamma)\vert_{\rm SM}$ and the two dashed lines delimit the
 interval~(\ref{allowed}).       }}
\label{branrat}
\end{figure}
We plot in fig.~\ref{branrat} the values obtained for the branching ratio
$\BR(\bsgamma)\vert_\MSSM$, as a function of $m$, for the points of
the parameter space $(m,M)$ remaining after the lower
limits~(\ref{expbounds}) are implemented. It is
understood in this section that the bounds~(\ref{expbounds}) are always
imposed and that the allowed points of the $(m,M)$ parameter space,
for fixed $\tb$ are only the dark points of
fig.~\ref{parspace}. The
degeneracy observed for fixed $m$, in fig.~\ref{branrat} corresponds to
the dependence on $M$. The
horizontal straight line indicates $\BR(\bsgamma)\vert_{\rm SM}$, i.e.
the value taken by the branching ratio in the SM for the
same values of parameters used to calculate
$\BR(\bsgamma)\vert_\MSSM$. We observe that, in spite of the
different experimental cuts on the supersymmetric masses here applied,
the result obtained for
$\tb=3$ reproduces quite closely the shape of points already
obtained in ref.~\cite{US} for $m_t=130\GeV$ and $\tb=2$,
while large enhancements are obtained for
larger values of $\tb$.

\begin{figure}[tb]
\begin{center}
\epsfxsize=10.0cm
\leavevmode
\epsfbox[30 179 555 765]{cnampl03.ps}
\end{center}
\caption[f6]{\small{Ratios of the supersymmetric contributions to the
 total amplitude for the decay $\bsgamma$ over the SM amplitude, as
 defined in~(\ref{ratios}), for $m_t=150\GeV$ and $\tb = 3$.
                }}
\label{amp03}
\end{figure}
In order to understand the origin of these enhancements,
we consider the ratios of the different supersymmetric contributions
to the total amplitude over the SM amplitude, in the same
spirit as already done in ~\cite{US}. We call these ratios
$R_i$. They are defined as:
\bea
\lefteqn{R_i \ \equiv \
\frac{{\cal A}(\bsgamma)_{i\phantom{M}}}
      {{\cal A}(\bsgamma)_{\rm SM}}  \ = \     }          \nn \\
& & = \frac{\eta^{-16/23}
            \left\{ -\frac{1}{2}\, {\cal A}^\gamma_i \, / \,
             \frac{e}{16 \pi^2}\, \right\} }
         {\eta^{-16/23}
            \left\{ -\frac{1}{2}\, {\cal A}^\gamma_{\rm SM} \, / \,
             \frac{e}{16 \pi^2}\,
           - \left[
          \frac{58}{135} \left( \eta^{10/23} - 1 \right) \, +
          \frac{29}{189} \left( \eta^{28/23} -1 \right)  \,
             \right]\, C_2(M_W)
            \right\} }\,
\label{ratios}
\eea
where the subindex $i$ indicates
$H^-$,~$\wti \chi^-$,~$\wti g$,~$\wti \chi^o$. The multiplicative
factor $\eta ^{-16/23}$ in the numerator has been omitted in the
neutralino contribution, as in ref.~\cite{US}.
\begin{figure}[tb]
\begin{center}
\epsfxsize=10.0cm
\leavevmode
\epsfbox[30 179 555 765]{cnampl15.ps}
\end{center}
\caption[f6]{\small{Same as in fig.~\ref{amp03} for $m_t=150\GeV$ and
 $\tb = 15$.                            }}
\label{amp15}
\end{figure}
We explicitly show in figs.~\ref{amp03},~\ref{amp15},~\ref{amp30}
the results obtained for $\tb = 3$,~$15$ and~$30$. For 
the remaining values of $\tb$,
$\tb = 9$,~$20$,~$25$, we obtain ratios $R_i$ which are
smooth transitions between the ratios plotted for two adjacent values
of $\tb$. We plot each $R_i$
versus one of the two types of masses relevant
for the particular contribution considered. As before, the vertical
width correspond to the degree of freedom still remaining when
$\tb$ and one other supersymmetric parameter are fixed.
\begin{figure}[tb]
\begin{center}
\epsfxsize=10.0cm
\leavevmode
\epsfbox[30 179 555 765]{cnampl30.ps}
\end{center}
\caption[f6]{\small{Same as in fig.~\ref{amp03} for $m_t=150\GeV$ and
 $\tb = 30$.                            }}
\label{amp30}
\end{figure}

Fig.~\ref{amp03} shows features analogous to those
obtained in~\cite{US} for $m_t=130\GeV$,~$\tb=2$. The largest
possible supersymmetric contribution, in this case, is indeed given
by the Higgs
contribution and the sum of the different contributions leads to an
overall, modest enhancement of $\BR(\bsgamma)\vert_\MSSM$ over
$\BR(\bsgamma)\vert_{\rm SM}$. The ratio $R_{H^-}$ is always positive
and, as shown in figs.~\ref{amp03},~\ref{amp15}~and~\ref{amp30},
is almost insensitive to the increase of $\tb$, for
$\tb > 3$. The only effect of $\tb$ on $R_{H^-}$ is an
indirect one: since larger values of $\tb$ provide in general lighter
$H^-$ in the same 2-dimensional parameter space $(m,M)$, one observe
that the smooth curve for $R_{H^-}$ stops earlier and earlier
when $\tb$ increases.

Positive and negative values of $R_i$ are obtained for all the
remaining supersymmetric contributions. The new feature shown by
figs.~\ref{amp15}~and~\ref{amp30} is the overwhelming predominance
of the chargino contribution, already ranging between -0.8 and 1
for $\tb = 9$ and exceeding the Higgs contribution by one
order of magnitude for $\tb$ between 20 and 25.
Nevertheless, as it is shown by the well spread distribution of
$R_{{\wti \chi}^-}$ above and below zero, it should be clear that
there exist regions of the parameter space $(m,M)$ where it is
$\vert R_{{\wti \chi}^-}\vert < R_{H^-} $ even for large values
of $\tb$, as well as regions where
$\vert R_{{\wti \chi}^-}\vert$ exceeds  $R_{H^-}$ for $\tb=3$.
The rapid increase of $ \vert R_{{\wti \chi}^-} \vert $
for increasing $\tb$ is due to the fact that
light enough ${\wti u}_1$ can be found, in these
cases, in coincidence with small values of $m_{{\wti \chi_2}^-}$,
as it can be seen by comparing figs.~\ref{cnhiggs}~and~\ref{cnstop}.
In particular, quite visible in these two figures are regions where
$m_{{\wti \chi_2}^-} $ and $m_{{\wti u}_1} $ have the minimum
allowed value of $45\GeV$.

A more modest, but still sizable increase is observed for the ratio
$\vert R_{\wti g} \vert$. This is due, in average, to
lighter squarks ${{\wti d_1}}$ (see fig.~\ref{cnstop}) and to larger
mixing parameters
${\wti g}$-${\wti d_i}$-${\wti d_j}~(i\neq j)$
obtained for larger values of $\tb$.
The same features, lighter ${{\wti d_1}}$ and larger
mixings ${\wti \chi_1}^0$-${\wti d_i}$-${\wti d_j}~(i\neq j)$,
together with the possibility of very light
neutralinos ($\sim 20\GeV$) are responsible for a quite rapid
increase of the neutralino
contribution. This increase is more reminiscent of the growth of the
chargino contribution (about a factor $20$ when going from
$\tb = 3$ to $\tb=30$) than the growth of the gluino
contribution (only a factor 5) and is clearly related to the
possibility of having small $\vert \mu_R\vert$ for large values of
$\tb$. It is not big enough, however, to beat the initial
disadvantage of this contribution and to make it reach
the value of Higgs and SM contributions.

One last comment has to be made on the relative sign of the different
ratios $R_i$. While $R_H$ is always positive, the signs of
$R_{{\wti \chi}^-}$, $R_{\wti g}$ and $R_{{\wti \chi}^0}$ for each
point of the supersymmetric parameter space depend on several
elements, such as the sign of $M$ and $\mu$, the relative size and
sign of gaugino and higgsino components for chargino and neutralino
contributions, and of the two components mediating for each
contribution the helicity flip of the process. Due to the complex
interplay among these elements, it is clear that all contributions,
even if subleading, can become important restrictions on the allowed
values of $\BR(\bsgamma)\vert_\MSSM$ are imposed. Only the
neutralino contribution plays a minor r\^ole in the determination of
the consequences of such a restriction.

Before starting this discussion, few more points have to be
clarified. The possibility of big enhancement-/suppression-factors
for $\BR(\bsgamma)$ in the case of relatively large values of $\tan\beta$
should alert one to check if similar results occur for other rare
processes. In this circumstance, one should worry about the possibility
that the regions of supersymmetric parameter space responsible for the
strongest deviations from the predictions for
$\BR(\bsgamma)\vert_{\rm SM}$ are not already excluded.

The decay modes
$b\to s l^+ l^-$, $b\to s \nu \bar{\nu}$,
$B_s \to \tau^+ \tau^-$, studied in \cite{BBOOK} and \cite{US}
are still far enough from being experimentally detected to pose
a real threat.
Surprises may in principle come from $B^0-\bar B^0$ oscillations. We
have explicitly checked that the supersymmetric contributions to the
$B^0_d-\bar B^0_d$ mixing, for $m_t=150\GeV$, yield deviations from
the SM prediction of about $10 \%$ and at most $15 \%$ in very limited
regions of the supersymmetric parameter space. These results confirm
even for large values of $\tb$ what was already
observed in \cite{US} for $m_t=130\GeV$ and $\tan\beta=2,8$. They
also confirm other claims made in the past, i.e. that supersymmetric
contributions to FCNC box diagrams yield results smaller than those
obtained in the SM, even for values of supersymmetric masses and
couplings which allow sizable enhancements of the
decay $\bsgamma$~\cite{USOLD}.

Given the results obtained for $\bsgamma$, one would expect even a more
spectacular supersymmetric enhancements for the decay $b\to s g $,
with an on-shell gluon. This decay mode contributes together with
$b \to s q \bar{q}$ and $b \to s g g $ to the so-indicated
$b \to s ``g''$, which leads to the observation of non-leptonic
charmless $B$-decays.
One would need enhancements of about two order of magnitude to have
$\Gamma(b \to s g)$ comparable to the SM value of
$\Gamma(b \to s q \bar{q})$. Thanks to the presence of the additional
coupling ${\wti g}$-${\wti g}$-$g$, this requirement may not be so hard
to be meet, especially for the large values of $\tb$ considered
here. In addition, since (smaller) enhancements to the decay mode
$b \to s q \bar{q}$ are also possible, one may think that deviations
from the SM predictions for non-leptonic, charmless $B$-decays could
be not completely negligible. A calculation which consistently
puts together all these contributions, however, is not available, as
yet. Moreover, any comparison with experimental results would have to
rely on the use of some specific hadronization model.

We leave this check aside, but we do consider a related issue, i.e.
the possibility that large enhancements of the decay $b \to s g$ may
bring substantial differences in $\BR(\bsgamma)\vert_\MSSM$ if the
form (\ref{corr2}) of the coefficient $C_7(\mu)$ is used. We obtain
changes in the patterns of points shown in the previous figures
never exceeding the 10\% level, even in the regions of parameter space
where the most optimistic enhancements of $b \to s g$ are found.

\subsection{Exclusion Patterns}

We come now to study the consequences that the restriction
(\ref{allowed}) to the allowed values of
$\BR(\bsgamma)\vert_\MSSM$ enforces on the supersymmetric
parameter space. Given the wide spread of points above and below the
central value for $\BR(\bsgamma)\vert_{\rm SM}$ shown in
fig.~\ref{branrat}, one expects these restrictions to be rather
drastic and more severe for the largest values of $\tb$,
where the biggest enhancements are present.
Except for $\tb=3$,
where still about 50\% of the original points remain after
condition~(\ref{allowed}) is imposed, only fractions ranging
from 24\% to 18\% survive for $\tb \geq 15$.
Thus, the interesting question to be answered is which ranges
of supersymmetric masses are left in the points
of the original parameter space which pass the test~(\ref{allowed}).

\begin{figure}[p]
\begin{center}
\epsfxsize=16cm
\leavevmode
\epsfbox[18 140 534 732]{sph2h_red2h.ps}
\end{center}
\caption[f5]{\small{Values of $\BR(\bsgamma)\vert_{\rm 2HDM}$, versus
 $m_{H^-}$ for the points contained within the narrow strip delimited
 by dashed lines in fig.~\ref{branrat}. The central straight
 line correspond to $\BR(\bsgamma)\vert_{\rm SM}$ and the two
 long-dashed lines enclose the interval~(\ref{allowed}), as in
 fig.~\ref{branrat}.       }}
\label{brred2h2h}
\end{figure}

Before tackling this issue, we plot the values which
$\BR(\bsgamma)\vert_{\rm 2HDM}$ (implicitly defined
in~(\ref{brtwohiggs})) has in these remaining points, i.e. the
values which $\BR(\bsgamma)\vert_\MSSM$ would assume if all the
supersymmetric contributions would be neglected. The results,
plotted versus $m_{H^+}$, are shown in fig.~\ref{brred2h2h}. Only
in the case $\tb=3$ the range of points shown in this figure
has been reduced in order to avoid the extremely large values of
$m_{H^-}$ which would be reached (up to $900\GeV$). The curves
obtained (including the portion not shown for $\tb=3$) are
all above the SM result (the solid line) and cross the
region~(\ref{allowed}), delimited by dashed lines, for
$m_{H^-}$ between $450$ and $500\GeV$ depending on the particular
value of $\tb$. We observe that values
between $100$ and $150\GeV$ for $m_{H^-}$ are still allowed, while
they would give too large branching ratios
in a 2DHM not embedded in the proper
supersymmetric scenario.

Thus, the possibility for the decay $t \to b H^+$ is still
open, although all the points of the $(m,M)$-parameter space where
$\Delta C_7^\prime (M_W),\,\Delta \BR^\prime\sim 0$ and where $H^\pm$ is
not prohibitively heavy are to be excluded.
This possibility is not restricted to the value
$m_t=150\GeV$, to which the previous figures refer, but it is, in
general, shared by all the reasonably large values of
$m_t$ allowed by the model.
For comparison, we show in fig.~\ref{rsthb} what remains of the
phase space shown in fig.~\ref{cnthb} after the condition~(\ref{allowed})
is imposed. We remind that this figure, together with fig.~\ref{cnthb}
is obtained by scanning the $(m,M)$-plane with a much finer grid than
the one used for all the remaining plots. This explain why the points
shown here for small values of $\tb$, typically $\tb=3$,
although rather sparse, seem nevertheless more numerous than those
shown in fig.~\ref{brred2h2h}. Wider ranges of $m_{H^-}$
remain for the large values of $\tb$ where stronger are the
restrictions of the parameter space $(m,M)$, since
the light ${H^-}$ are, in this case, found in wider
portions of the plane $(m,M)$.

\begin{figure}[h]
\begin{center}
\epsfxsize=12.4cm
\leavevmode
\epsfbox[60 360 523 751]{thb150_smrs.ps}
\end{center}
\caption[f9]{\small{What remains of the values of $m_{H^+}$ shown in
  fig.~\ref{cnthb} after excluding all points of the 2-dimensional
  parameter space $(m,M)$ which lead to
  $\BR(\bsgamma)\vert_\MSSM$ outside the interval~(\ref{allowed})%
.                }}
\label{rsthb}
\end{figure}

Since the chargino contribution is by far the dominant one,
in large regions of the supersymmetric parameter space here
considered, one should focus
on this contribution and check whether sizable regions of the
parameter space explorable at LEP~II are not already excluded
by~(\ref{allowed}). To this aim,
we plot the points of fig.~\ref{branrat} in the plane
$(\mu_R,M_2)$ as shown in fig.~\ref{lepexcl}. We distinguish the
regions allowed and forbidden by~(\ref{allowed}) by making use of
faint and thick points, respectively. As can be seen by comparing
this figure with fig.~\ref{mumtwo}, the white areas denote the regions
where no realizations of the MSSM are possible and the regions
excluded by the limits~(\ref{expbounds}).
\begin{figure}[p]
\begin{center}
\epsfxsize=16cm
\leavevmode
\epsfbox[18 140 534 732]{lepchi.one}
\end{center}
\caption[f5]{\small{Possible realizations of the MSSM,
  compatible with the lower bounds~(\ref{expbounds})
  plotted in the $(\mu_R,M_2)$-plane (compare with
  fig.~\ref{mumtwo}). The dark areas
  correspond to the region inside the narrow strip delimited by
  dashed lines in fig.~\ref{branrat}, the faint dots to the region
  outside it. Superimposed are the contour lines
  $m_{{\wti \chi_2}^-} = 45\GeV$ (dashed lines) and
  $m_{{\wti \chi_2}^-} = 90\GeV$ (solid lines).       }}
\label{lepexcl}
\end{figure}
\begin{figure}[p]
\begin{center}
\epsfxsize=16cm
\leavevmode
\epsfbox[18 140 534 732]{chgl.ps}
\end{center}
\caption[f5]{\small{Points of the parameter space $(m,M)$
 compatible with the lower bounds~(\ref{expbounds}),
 plotted in the plane
 $(m_{{\wti \chi_2}^-},m_{\wti g})$. The thick and faint dots
 correspond to the regions allowed and forbidden by
 condition~(\ref{allowed}).
                 }}
\label{chglexcl}
\end{figure}

\begin{figure}[p]
\begin{center}
\epsfxsize=16cm
\leavevmode
\epsfbox[18 140 534 732]{stopgl.ps}
\end{center}
\caption[f5]{\small{Same points of fig.~\ref{chglexcl}
 plotted in the plane
 $(m_{{\wti u}_1},m_{\wti g})$. The thick and faint dots
 correspond to the regions allowed and forbidden by
 condition~(\ref{allowed}).           }}
\label{stopglexcl}
\end{figure}
Superimposed are the contour lines
$m_{{\wti \chi_2}^-} = 45\GeV$ (dashed lines) and
$m_{{\wti \chi_2}^-} = 90\GeV$ (solid lines), corresponding to
regions probed at LEP~I and explorable at LEP~II, respectively (see
for comparison~\cite{GRIVAZ}). We observe that the reduction of
parameter space is rather significant. We remind that not all the
physical points compatible with the lower limits~(\ref{expbounds})
are plotted in figure~\ref{lepexcl}. Points surviving
condition~(\ref{allowed}) in the branch with negative $M_2$ and
negative $\mu_R$ are present for values of $\mu_R$ not shown in this
figure. For all values of $\tb$ considered here, we always find
points where the lightest chargino can have masses compatible with
the experimental cut applied in this analysis,
$m_{{\wti \chi_2}^-} \sim 45 \GeV$, among the points remaining after
imposing condition~(\ref{allowed}). This can be more explicitly seen
in fig.~\ref{chglexcl} where the same points are plotted in the plane
$(m_{{\wti \chi_2}^-},m_{\wti g})$. Although the regions excluded by
condition~(\ref{allowed}) are quite wide, the chargino searches at
LEP~II do not appear to be threatened by a measurement of $\bsgamma$
at the SM level.

\begin{figure}[tb]
\begin{center}
\epsfxsize=10.0cm
\leavevmode
\epsfbox[30 179 555 765]{ampls30_smrs.ps}
\end{center}
\caption[f6]{\small{Points of fig.~\ref{amp30} surviving
 condition~(\ref{allowed}).                 }}
\label{amp30smrs}
\end{figure}

Since the situation
$\Delta \BR^\prime >> \BR(\bsgamma)\vert_{\rm SM,2HDM}$, due to
large chargino contributions, is observed in non-trivial portions of
the supersymmetric parameter space, there remain the possibility that
the lightest eigenvalue of the up-squark mass matrix ${\wti u}_1$,
the stop mass, is the mass more seriously affected by the
restriction~(\ref{allowed}). To explore this possibility, we plot the
points of fig.~\ref{branrat} in the plane
$(m_{{\wti u}_1},m_{\wti g})$, as shown in
fig.~\ref{stopglexcl}. Again, the thick and faint points denote the
regions allowed and forbidden by~(\ref{allowed}). We observe that the
values of $m_{\wti u_1}$ in the ``allowed'' points are never below
$70-80\GeV$. The regions of parameter space where
$m_{\wti u_1}$ and $m_{{\wti \chi }_2^-} $ are simultaneously very
small ($\sim 45\GeV$), responsible for the conspicuous enhancements
observed, are removed by condition~(\ref{allowed}). Given the
distribution of masses observed in
figs.~\ref{cncharg}~and~\ref{cnstop},
however, light enough ${{\wti \chi_2}^-}$
can still be present in coincidence with quite heavy ${\wti u}_1$.

Only discrete values of $\tb$ have been considered here. Therefore,
the possibility of light ${\wti u}_1$ escaping our search still
exists, in principle. Nevertheless, since our choice of values should give
us an almost complete overview on this parameter, we consider this
possibility rather remote.

Summarizing, all points of the $(m,M)$-plane where
$\Delta \BR \sim 0$, corresponding to the situation
$\Delta C_7(M_W) \sim 0$, the pure 2HDM case, or
$\Delta C_7(M_W) \sim-2 C_{\rm 2HDM}$, as well as the points with
$\Delta \BR >> \BR(\bsgamma)\vert_{\rm SM,2HDM}$ have to be
excluded. The possibility that substantial regions of the
supersymmetric parameter space remain, relies on the strong
destructive interferences among {\it all} supersymmetric
contributions, and on the rather large uncertainty of the SM
prediction.

The typical situation observed in the regions of parameter space
surviving condition~(\ref{allowed}) is that negative contributions
$R_{{\wti \chi}^-} + R_{\wti g}$ are obtained, matching in size
$R_H$. The importance of all supersymmetric contributions, in
particular even of the more modest gluino contribution, is shown
explicitly in fig.~\ref{amp30smrs} for the case $\tb = 30$. We plot in
this figure the ratios $R_i$ for the points remaining after
condition~(\ref{allowed}) is imposed. We observe how significant a role
the small value $R_{\wti g}\sim .2$ plays for the survival of a region
where the chargino contribution is still rather large,
$R_{{\wti \chi }^-}\sim -2.5$. Similar interesting interplays of
different contributions are observed for
all values of $\tb$ considered. Only the neutralino contribution
has little influence on the exclusion patterns induced
by~(\ref{allowed}).

Values of $\as(M_W)$ different from the one used for the determination
of the interval~(\ref{allowed}) are not likely to change the
qualitative features observed here. In contrast, conditions more
restrictive than~(\ref{allowed}) may exclude wider regions of the
supersymmetric parameter space. These conditions, though, should be
imposed on calculations which could claim a precision higher than the
one so far achieved.

\newpage

\section{Conclusions}

We have analyzed the supersymmetric contribution to
$\bsgamma$ within the framework of the MSSM with radiative breaking
of $SU(2)\times U(1)$. Wide regions of supersymmetric
parameter space have been surveyed. Results have been explicitly
reported only for $m_t=150\,$GeV and different values of $\tb$, in a
range wide enough to allow a fairly complete overview of the
situations one encounters when considering ratios of top and bottom
yukawa couplings from $h_t/h_b>>1$ to $h_t/h_b \sim 1$.

The modest enhancements over the SM prediction obtained for relatively
large ratios $h_t/h_b$ ($m_t = 130\,$GeV, $\tb = 2,6$) in
ref.~\cite{US} are confirmed, whereas substantial
deviations from the SM prediction are found for large values of $\tb$,
due, primarily, to conspicuous chargino contributions.

As pointed out by several authors,
a measurement of the branching ratio for this decay at the SM level
can exclude relevant portions of the supersymmetric parameter space.
The still big uncertainty plaguing the theoretical determination
of $\BR(\bsgamma)\vert_{\rm SM}$, however, and destructive
interferences among the different supersymmetric contributions almost
produce complicated exclusion patterns. The overall effect is that
no clear-cuts bounds on chargino and charged Higgs masses
are obtained, whereas an increase in the degeneracy of the
up-squark mass matrix, is enough to
reduce the large chargino contributions and bring
$\BR(\bsgamma)\vert_{\rm MSSM}$ down to the
accepted values~(\ref{allowed}).

Stronger constraints could be presumably obtained if the error on the
theoretical SM and MSSM prediction would be reduced. To begin with,
some effort should be put in refining the level of precision of
the SM prediction, by calculating the NLO QCD correction to the
decay rate. Were more accurate SM results 
available, one would have to worry
about matching the precision of the MSSM calculation to the SM level.

Besides the complex evaluation of NLO QCD corrections within the MSSM
framework, one should also attempt a refinement of the determinations
of the physical realizations of the MSSM. The first, obvious
improvement with respect to the calculation here presented is the
inclusion of radiative corrections to the tree-level Higgs
potential. Moreover, the underlying grand-unification structure of the
MSSM model should be better specified. Issues related to the
yukawa couplings unification (as for example the accuracy at which
these unification conditions have to be imposed) are to be
settled. Needed are also evaluations of the uncertainties induced on
the low-energy supersymmetric mass spectrum by approximations made
while running down from the Planck scale $M_P$ to the electroweak
scale $M_Z$, such as the neglect of the heavy particles required by
the grand-unification gauge group, as well as the neglect of
non-renormalizable terms still sizable at the Planck scale.
Finally, once more control on these aspects is reached, and more
indications about the underlying grand-unified structure are obtained,
the constraints coming from the lack of observation of nucleon decays
should also be implemented.

\vskip 3ex
\noindent
{\it Note Added:} After the completion of this work we became aware
of the existence of refs.~\cite{LAST,OKADA}. The first one confirms
the results of ref.~\cite{OSHIMO} for the chargino contribution,
points to the light stop mass as to the culprit of the large
enhancements obtained for large values of $\tb$ and as to the
parameter likely to be affected by
measurements of the decay $\bsgamma$ at the SM level.
Different conclusions are reached in~\cite{OKADA}, where it is argued
that very light stops are still possible, in contrast with the results
of ref.~\cite{LAST} and the ones obtained in the present analysis.

\vskip 3ex
\noindent
{\bf Acknowledgements}
The author acknowledges discussions with A.~Ali, J.F.~Grivaz,
H.~Haber, R.~Hempfling, A.~Masiero, M.~Peskin, S.~Pokorski and
P.~Zerwas. She is also grateful to C. Vohwinkel for tips on handling
and storing the numerous and huge data files needed for this
analysis.

\appendix
\noindent{\Large \bf Appendix:  Errata}

In this appendix we list errata for refs.~\cite{US}~and~\cite{IO}.

\vspace{0.5cm}
\noindent{\bf ref.~\cite{US}:}

$\bullet$
The symbols $H_1^0$, $H_2^0$, reserved for the two CP-even neutral
Higgs fields, should be replaced in eq.~(8) by $H_1^1$, $H_2^2$,
respectively the first component of doublet $H_1$ and the second
component of doublet $H_2$. Following the notation of
ref.~\cite{GUNHAB}, adopted in~(\cite{US}) as well as in the present
paper, these two components are related to the physical fields
$H_1^0$, $H_2^0$ and $H_3^0$ according to:
\bea
H_1^1 & = & v_1 +\frac{1}{\sqrt{2}} \left(
             H_1^0 \cos\alpha -H_2^0 \sin\alpha + i H_3^0 \sin \beta
                                    \right)              \nn \\
H_2^2 & = & v_2 +\frac{1}{\sqrt{2}} \left(
             H_1^0 \sin\alpha +H_2^0 \cos\alpha + i H_3^0 \cos \beta
                                    \right)\,.           \nn
\eea

$\bullet$
The equation (A.1) in Appendix~A should read
\bea
\frac{\d m_Q^2}{\d t} &=&
    \left(\frac{16}{3}\wti\alpha_3 M_3^2
    + 3\wti\alpha_2 M_2^2
    + \frac{1}{9}\wti\alpha_1 M_1^2\right)
    {\mathop{\bf 1}}                            \nn      \\
  & & - \ \frac{1}{2} \left [
    \wti Y_U\wti Y_U^\dagger m_Q^2
    + m_Q^2\wti Y_U\wti Y_U^\dagger
    + 2 \left( \wti Y_U m_U^2 \wti Y_U^\dagger
    + \bar\mu_2^2 \wti Y_U\wti Y_U^\dagger
    + m^2 \wti Y_U^A\wti Y_U^{A\dagger}
    \right) \right ]                            \nn      \\
  & & - \ \frac{1}{2} \left [
    \wti Y_D\wti Y_D^\dagger m_Q^2
    + m_Q^2\wti Y_D\wti Y_D^\dagger
    +2 \left( \wti Y_D m_D^2 \wti Y_D^\dagger
    + \bar\mu_1^2 \wti Y_D\wti Y_D^\dagger
    + m^2 \wti Y_D^A\wti Y_D^{A\dagger}
    \right) \right ]\,. \hspace{1.5em}\null     \nn
\eea
The spurious factor 2 multiplying the term $(1/9){\wti{\alpha}_1}M_1^2$
in the first bracket on the left-hand side of (A.1) is clearly a
misprint 
and was not included in the RG equations used in the actual calculations.

\vspace{0.5cm}
\noindent{\bf ref.~\cite{IO}:}

$\bullet$
The $2\times 2$ mass matrices in eqs.~(54) and~(55) should read:
\[
M_{\wti t}^2   =   \left( \begin{array}{cc}
 m_{\wti{Q}_{33}}^2 +m_t^2  -\vert DT_U^L\vert \ \            &
       \left(A_t m +\mu_{\rm R} \cot \beta \right) m_t   \nn  \\[1.5ex]
\!\! \left(A_t m + \mu_{\rm R} \cot \beta \right) m_t         &
 \ m_{\wti{U}_{33}}^2 +m_t^2  -\vert DT_U^R\vert \
                    \end{array} \right)\phantom{\ .}            \nn
\label{stopmatrix}
\]
\[
M_{\wti b}^2   =  \left( \begin{array}{cc}
 m_{\wti{Q}_{33}}^2 +m_b^2  +\vert DT_D^L\vert \ \           &
     \left(A_t m +\mu_{\rm R} \ \tb \right) m_b   \nn \\[1.5ex]
\!\! \left(A_t m +\mu_{\rm R} \ \tb \right) m_b       &
 \ m_{\wti{D}_{33}}^2 +m_b^2  +\vert DT_D^R\vert \
                  \end{array} \right) \ .                      \nn
\label{sbotmatrix}
\]

\newpage




\begin{thebibliography}{99}
\bibitem{CLEO}
 E.~Thorndike et al., CLEO Collaboration, CLEO preprint
 CLN 93/1212, CLEO 93-06.

\bibitem{QCD}
 S.~Bertolini, F.~Borzumati and A.~Masiero, \prl{59}{87}{180}; \\
 N.G.~Deshpande, P.~Lo, J.~Trampetic, G.~Eilam and P.~Singer,
                          \prl{59}{87}{183}.

\bibitem{GRSPWI}
 B.~Grinstein, R.~Springer and M.B.~Wise, \plb{202}{88}{138};
                                          \npb{339}{90}{269}.

\bibitem{OTHERS}
 R.~Grinjanis, P.J.~O'Donnell, M.~Sutherland and H.~Navelet,
             \plb{213}{88}{355},  (E)~\ib{286}{92}{413}; \\
 G.~Cella, G.~Curci, G.~Ricciardi and A.~Vicer\'e,
  \plb{248}{90}{181};                                         \\
 A.~Ali and C.~Greub, \plb{259}{91}{182}; \zpc{49}{91}{431}.

\bibitem{GRICHO}
 P.~Cho and B.~Grinstein, \npb{365}{91}{279}.

\bibitem{MISIAK}
 M.~Misiak, \plb{269}{91}{161} and \npb{393}{93}{23}.

\bibitem{ALIGR}
 A.~Ali and C.~Greub, DESY preprint DESY 93-065, ZU-TH 11/93, (1993).

\bibitem{HEWETT}
 J.L.~Hewett, \prl{70}{93}{1045}.

\bibitem{BARGER}
 V. Barger, M.S.~Berger and R.J.~Phillips, \prl{70}{93}{1368}.

\bibitem{DIAZ}
 M.A.~Diaz, \plb{304}{93}{278}.

\bibitem{OSHIMO}
 N.~Oshimo, \npb{404}{93}{20}.

\bibitem{TANIMOTO}
 T. Hayashi, M. Matsuda and M. Tanimoto, Kogakkan University preprint,
  KU-01-93, AUE-01-93, EHU-01-93 (1993).

\bibitem{BARGIUD}
 R.~Barbieri and G.F.~Giudice,~\plb{309}{93}{86}.

\bibitem{NANOPOU}
J.L.~Lopez, D.~Nanopoulos, G.T.~Park, \prd{48}{93}{974}.

\bibitem{US}
 S.~Bertolini, F.~Borzumati, A.~Masiero and
                               G.~Ridolfi,~\npb{353}{91}{591}.

\bibitem{BBOOK}
S.~Bertolini, F.~Borzumati and A.~Masiero in
   {\it ``B Decays''}, S. Stone~(ed.), World Scientific
   (Singapore), 1992.

\bibitem{IO}
  F.~Borzumati, in {\it ``Phenomenological Aspects of Supersymmetry''},
   W. Hollik, R. R\"uckl and J. Wess~(eds.), Springer Lectures Notes
   in Physics, Springer Verlag (Heidelberg), 1992.

\bibitem{LANGPOL}
 P.~Langacker and N.~Polonsky, \prd{47}{93}{4028}.

\bibitem{GUNHABSHER}
 J.F. Gunion, H.E. Haber and M. Sher, \npb{306}{88}{1}.

\bibitem{ALPHAS}
 S.~Bethke, Heidelberg University preprint, HD-PY-92-13,
 (1992), Talk given at 26th International Conference on High Energy
 Physics (ICHEP 92), Dallas, TX, 6-12 Aug 1992.

\bibitem{PDB}
 Particle Data Group, {\it Phys. Rev.}~45 (1993) Part~II.

\bibitem{YUKUNIF}
 V.~Barger, M.S.~Berger and P.~Ohmann, \prd{47}{93}{1093};     \\
 M.~Carena, S.~Pokorski and C.E.M.~Wagner, Max-Planck Inst.
  preprint, MPI-Ph/93-10~(1993);                              \\
 P.~Langacker and N.~Polonsky, University of Pennsylvania preprint,
  UPR-0556-T~(1993).

\bibitem{OLECH}
 M.~Olechowski and S.~Pokorski, \npb{404}{93}{590}.

\bibitem{CDFBOUNDS}
 F.~Abe, et al., CDF Collaboration, \prl{69}{92}{3439}.

\bibitem{STOPBOUND}
 R.~Keraenen, DELPHI internal report, DELPHI 92-172 PHYS 255 (1993).

\bibitem{HIGGSBOUND}
 A. Sopczak, CERN-PPE/92-137 (1992).

\bibitem{USOLD}
 S.~Bertolini, F.~Borzumati and A.~Masiero, \plb{192}{87}{437}; \\
 S.~Bertolini, F.~Borzumati and A.~Masiero, \npb{294}{87}{321}; \\
 S.~Bertolini, F.~Borzumati and A.~Masiero,
  \plb{194}{87}{545}, (E)~\ib{198}{87}{590}.

\bibitem{GRIVAZ}
 J.-F.~Grivaz, in {\it ``Physics and Experiments with Linear Colliders,
 Saariselk\"a, Finland, 9-14 September 1991''}, vol.~I,
  R.~Orava, P.~Eerola and M.~Nordberg~(eds.),
  World Scientific (Singapore), 1992.

\bibitem{LAST}
 R. Garisto and J.N.~NG, TRIUMF preprint, TRI-PP-93-66 (1993).

\bibitem{OKADA}
 Y.~Okada, KEK Preprint 93-68, KEK-TH-365 (1993).

\bibitem{GUNHAB}
 J.F.~Gunion and H.E.~Haber, \npb{272}{86}{1}.

\end{thebibliography}
\end{document}